\documentclass[sigconf,nonacm]{acmart}

\usepackage{array}
\usepackage{xcolor}
\usepackage{caption}
\usepackage{listings}
\usepackage{booktabs} % For formal tables
\usepackage{graphicx} % For figure environment
\usepackage{multirow}
\usepackage{textcomp}
\usepackage{subcaption}
\usepackage{algorithm2e}
\usepackage{algpseudocode}
\usepackage{ulem}
\usepackage{hyperref}

\definecolor{codegreen}{rgb}{0,0.6,0}
\definecolor{codegray}{rgb}{0.5,0.5,0.5}
\definecolor{codepurple}{rgb}{0.58,0,0.82}
\definecolor{backcolour}{rgb}{0.95,0.95,0.92}

\lstdefinestyle{mystyle}{
    backgroundcolor=\color{backcolour},   
    commentstyle=\color{codegreen},
    keywordstyle=\color{magenta},
    numberstyle=\tiny\color{codegray},
    stringstyle=\color{codepurple},
    basicstyle=\footnotesize,
    breakatwhitespace=false,         
    breaklines=true,                 
    captionpos=b,                    
    keepspaces=true,                 
    numbers=left,                    
    numbersep=5pt,                  
    showspaces=false,                
    showstringspaces=false,
    showtabs=false,                  
    tabsize=2
}

\newcommand\blfootnote[1]{%
  \begingroup
  \renewcommand\thefootnote{}\footnote{#1}%
  \addtocounter{footnote}{-1}%
  \endgroup
}

\copyrightyear{2021}
\acmYear{2021}
\setcopyright{none}
\acmConference[Advanced Micro Devices, Inc. Technical Report]{Advanced Micro Devices, Inc. Technical Report}{December 14}

\setcitestyle{square}

\begin{document}

% Title. 
% If your title is long, consider \title[short title]{full title} - "short title" will be used for running heads.
\title{GI-1.0: A Fast Scalable Two-Level Radiance Caching Scheme for Real-Time Global Illumination}

% Authors.
\author{Guillaume Boissé}
\affiliation{\institution{\normalsize{Advanced Micro Devices, Inc.}}\country{France}}
\author{Sylvain Meunier}
\affiliation{\institution{\normalsize{Advanced Micro Devices, Inc.}}\country{France}}
\author{Heloise de Dinechin}
\affiliation{\institution{\normalsize{Advanced Micro Devices, Inc.}}\country{France}}
\author{Matthew Oliver}
\affiliation{\institution{\normalsize{Advanced Micro Devices, Inc.}}\country{Australia}}
\author{Pieterjan Bartels}
\affiliation{\institution{\normalsize{Advanced Micro Devices, Inc.}}\country{Belgium}}
\author{Alexander Veselov}
\affiliation{\institution{\normalsize{Advanced Micro Devices, Inc.}}\country{Germany}}
\author{Kenta Eto}
\affiliation{\institution{\normalsize{Advanced Micro Devices, Inc.}}\country{Japan}}
\author{Takahiro Harada}
\affiliation{\institution{\normalsize{Advanced Micro Devices, Inc.}}\country{USA}}

% \author{Brittany Rowland-Smith}
% \affiliation{%
%   \institution{St. Olaf College}}
% \email{br-s@gmail.com}

% \author{Nicholas Badeeri}
% \affiliation{%
%   \institution{MathWorks, Inc.}}
% \email{badeeri@mathworks.com}

% \author{Andrew Joseph Foyt}
% \affiliation{%
%   \department{College of Engineering}
%   \institution{University of Houston}}
% \email{foyt_aj@uh.edu}

% This command defines the author string for running heads.
% \renewcommand{\shortauthors}{DeJohnette, Rowland-Smith, Badeeri, and Foyt}
\renewcommand{\shortauthors}{Boissé et al.}

% abstract
\begin{abstract}
Real-time global illumination is key to enabling more dynamic and physically realistic worlds in performance-critical applications such as games or any other applications with real-time constraints.
Hardware-accelerated ray tracing in modern GPUs allows arbitrary intersection queries against the geometry, making it possible to evaluate indirect lighting entirely at runtime.
However, only a small number of rays can be traced at each pixel to maintain high framerates at ever-increasing image resolutions.

Existing solutions, such as probe-based techniques, approximate the irradiance signal at the cost of a few rays per frame but suffer from a lack of details and slow response times to changes in lighting.
On the other hand, reservoir-based resampling techniques capture much more details but typically suffer from poorer performance and increased amounts of noise, making them impractical for the current generation of hardware and gaming consoles.

To find a balance that achieves high lighting fidelity while maintaining a low runtime cost, we propose a solution that dynamically estimates global illumination without needing any content preprocessing, thus enabling easy integration into existing real-time rendering pipelines.

\blfootnote{\url{https://github.com/GPUOpen-LibrariesAndSDKs/Capsaicin}, email: \{ Guillaume.Boisse, Sylvain.Meunier, Heloise.Dupontdedinechin, Matthew.Oliver, Pieterjan.Bartels, Kenta.Eto, Takahiro.Harada \}@amd.com \\ Advanced Micro Devices, Inc. Technical Report No. 22-10-9831, October 18, 2022. }
\end{abstract}

%CCS
%keywords

% A "teaser" figure, centered below the title and authors and above the body of the work.
\begin{teaserfigure}
  \centering
      \begin{subfigure}[b]{0.498\textwidth}
        \centering
        \includegraphics[width=\textwidth]{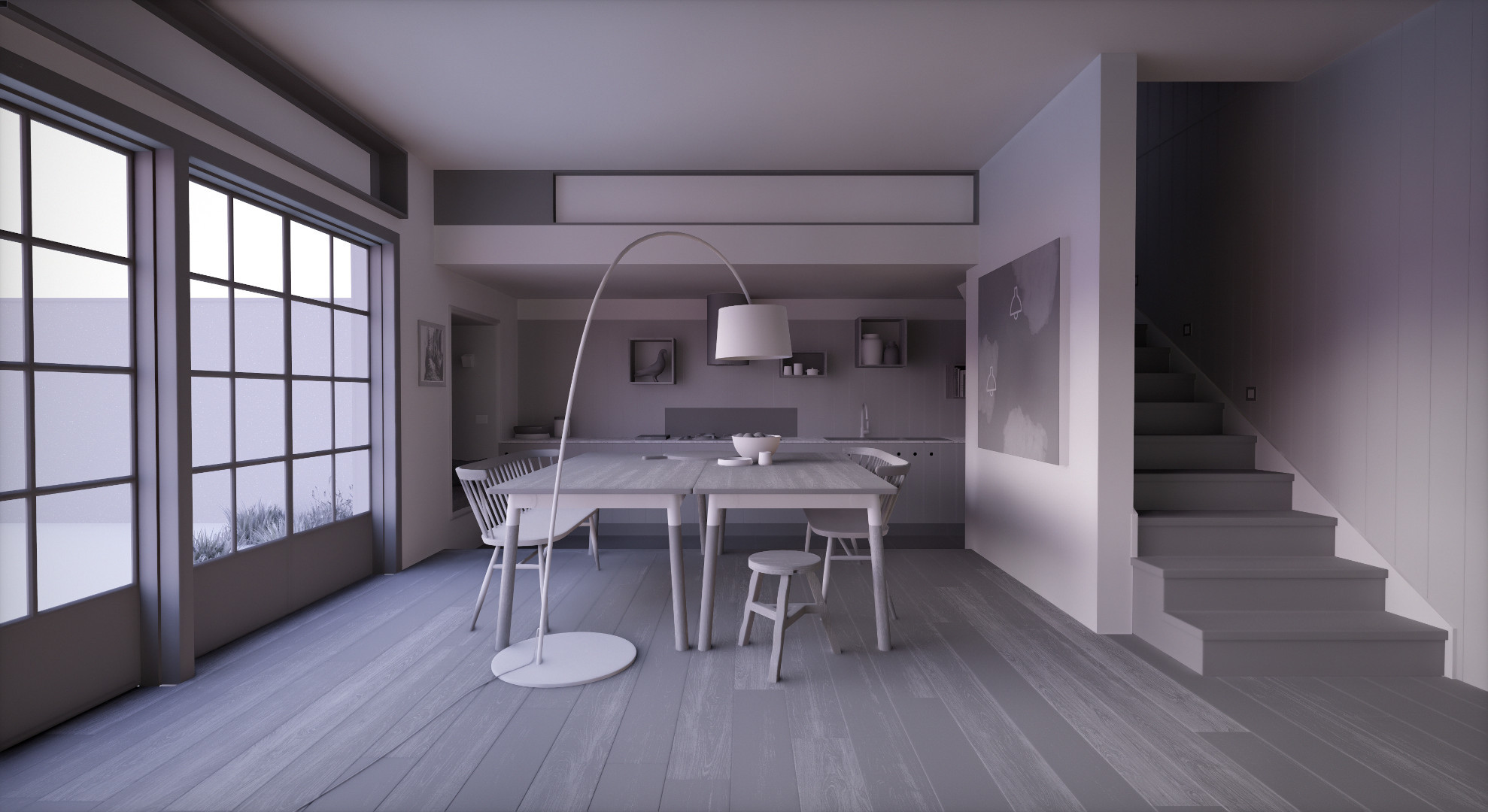}
    \end{subfigure}
    \hfill
      \begin{subfigure}[b]{0.498\textwidth}
        \centering
        \includegraphics[width=\textwidth]{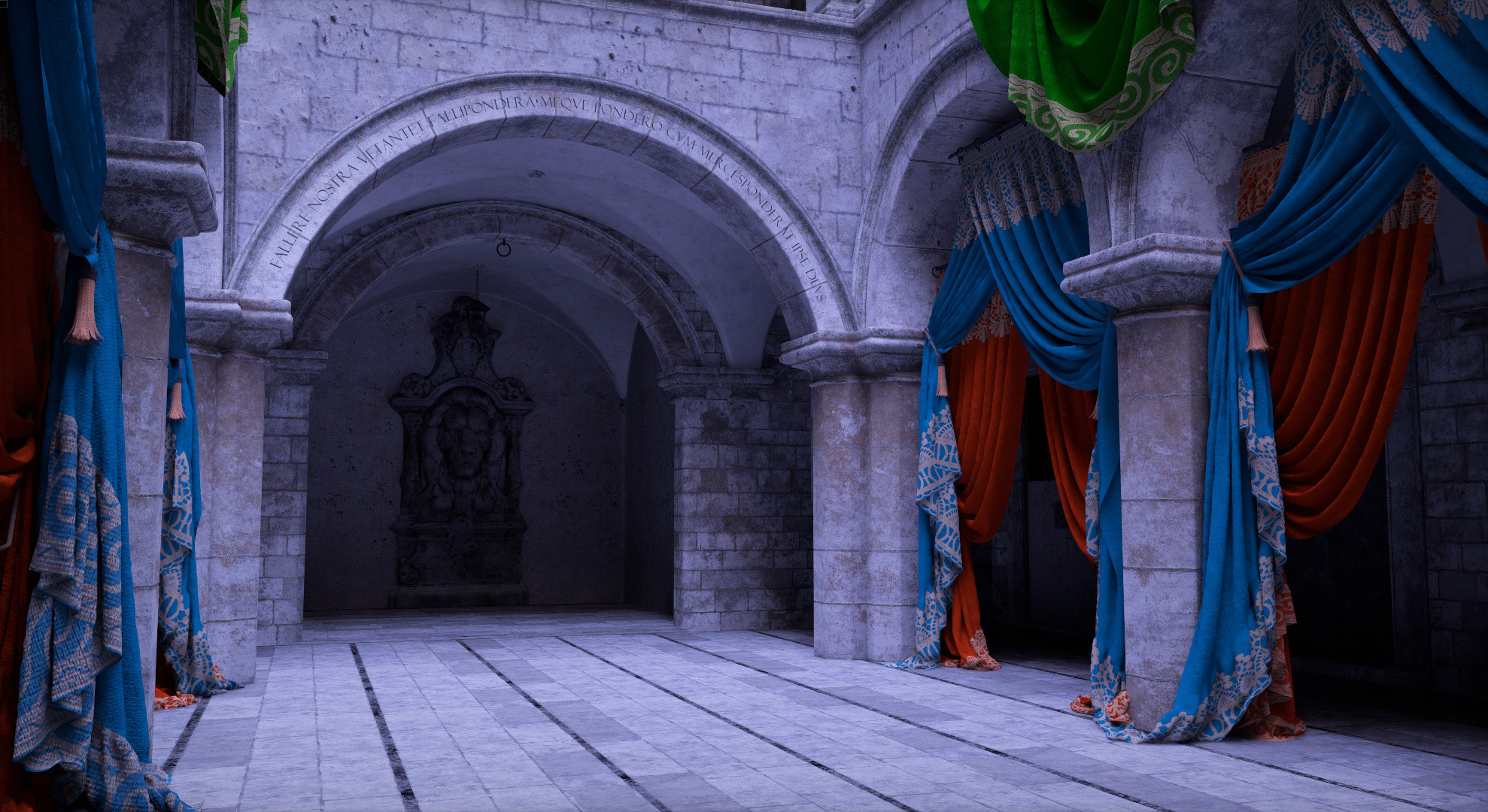}
    \end{subfigure}
  \caption{Kitchen and Sponza scenes rendered with direct and indirect lighting calculated using our GI-1.0 pipeline in 3.5ms and 4.2ms respectively at 1080p on Radeon\texttrademark\ RX 6900 XT.}
  \label{fig:teaser}
\end{teaserfigure}

\maketitle
\section{Introduction}

Probe-based techniques are often used in applications where a high framerate is required \cite{10.1109/38.656788}.
Light probes were precomputed in the past, but real-time hardware ray tracing makes it possible to compute them dynamically at runtime. 
Majercik {\it et al.} proposed Dynamic Diffuse Global Illumination to cache the irradiance field into a set of dynamically ray-traced probes organized in world-space grids \cite{Majercik2019Irradiance}.
The per-pixel irradiance value can then be estimated by interpolating from the eight neighboring probes, taking the visibility information into account in the form of a Chebychev inequality test \cite{10.1145/1111411.1111440}.
While this reduces the light leaking issue that plagued previous probe systems, the visuals often end up looking flat as the irradiance near occluders is typically of higher frequency than what the spatial resolution of the probe field can capture.
%Furthermore, such systems often exhibit a slow response to changes in lighting.

Techniques using resampled importance sampling have been actively explored recently \cite{10.2312:EGWR:EGSR05:139-146,  slc}.
Reservoir-based Spatiotemporal Importance Resampling (ReSTIR) heavily relies on reservoir resampling to get high-quality samples, which results in a higher quality sampling for direct illumination \cite{10.1145/3386569.3392481}.
It was further extended to indirect illumination \cite{10.1145/3478512.3488613}.  
ReSTIR Global Illumination (ReSTIR GI) \cite{10.1111:cgf.14378} and, more recently, ReSTIR Path Tracing (ReSTIR PT) \cite{10.1145/3528223.3530158} propose to trace rays per pixel and rely on reservoir-based resampling to efficiently share path sampling information across neighboring pixels and frames.
Such approaches yield promising visual results but are too expensive for performance-critical applications such as games or any other applications with real-time constraints.
Furthermore, tracing per pixel as opposed to interpolating between probes introduces significant amounts of noise in the image that must be filtered \cite{10.1145/3105762.3105770}.
Aggressive filtering of noisy signals can lead to a loss in quality and increased difficulty in keeping up with lighting changes.
Calculating a temporal gradient from the radiance signal \cite{10.1145/3233301} can help inform the denoising in such cases, but it is not trivial to compute, nor does it help when the input signal is overly noisy in the first place.

Screen Space Radiance Caching (SSRC) \cite{lumen} introduces a novel method that caches the radiance in probes spawned directly onto primary visible surfaces.
The approach has advantages similar to the probe system and shows little noise after interpolating the lighting for every pixel.
However, unlike world probes, screen probes do not suffer from light and occlusion leaks as they are always placed precisely on the geometry.
Furthermore, they offer a significantly higher density radiance representation, which leads to higher-fidelity visuals.

In this paper, we build on the SSRC approach and introduce several key contributions to improve performance and visual quality.
Most importantly, we introduce a novel caching algorithm to achieve temporally stable lighting without needing an additional world-space structure as required in \cite{lumen}.
We will show how we connect the screen probe rays to a secondary level of radiance caching based on spatial hashing \cite{cite:mppsf} to achieve fast, high-fidelity, and leak-free dynamic global illumination as shown in Figure \ref{fig:teaser}.

\section{GI-1.0}

For real-time purposes, only a few samples can be used for each pixel, even with today's high-end GPUs, to remain practical.
It is, therefore, essential to try and ensure that most, if not all, samples contribute to the lighting estimate in a meaningful way to keep the variance to a minimum.
Indeed, any ray or path not hitting a light source can potentially severely impact the quality of the rendered animation by introducing noise that would require excessive amounts of filtering.

\begin{figure}
  \centering
  \includegraphics[width=0.8\linewidth]{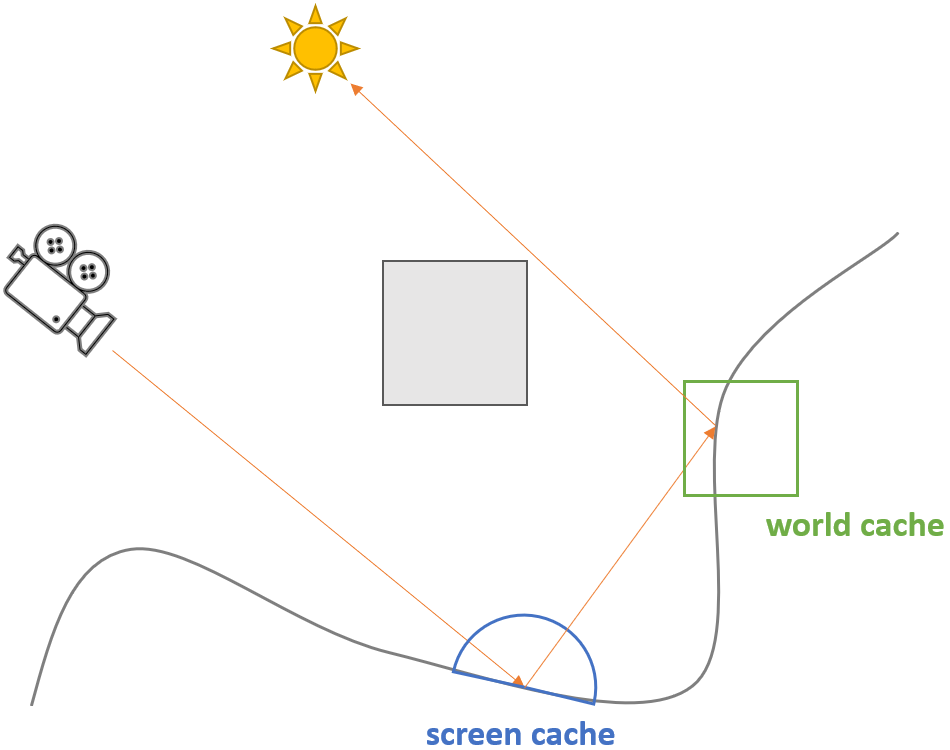}
  \caption{GI-1.0 two-level radiance caching scheme.}
  \label{fig:2-level_radiance-caching}
\end{figure}

We design our global illumination pipeline to make the most of every sample by reusing the lighting information across space and time for both sampling and filtering.
We do so by persisting the scene illumination inside two distinct levels of radiance caching, as illustrated in Figure \ref{fig:2-level_radiance-caching}:

\begin{itemize}
  \item The \emph{screen cache} caches the incoming radiance for primary path vertices inside probes placed directly onto primary visible surfaces, and offers a detailed lighting representation thanks to a large number of probes.
  \item The \emph{world cache} caches the outgoing radiance for secondary path vertices and, despite being less detailed than the screen cache, has the advantage of being stable and persistent.
\end{itemize}

We will demonstrate how this setup allows to compute high-fidelity and temporally responsive direct and indirect lighting using sampling rates as low as \textonequarter{} sample per pixel.

\subsection{Screen Cache}

This section describes how screen probes are spawned sparsely directly onto pixels, how we manage the integrity of the caching data structure across frames, and how we adapt the filtering heuristics based on depth to ensure temporally stable lighting at any distance.

Similarly to \cite{lumen}, we encode the incoming radiance across the oriented hemisphere into an 8x8 atlas using an octahedral projection mapping \cite{Cigolle2014Vector}.

\subsubsection{Temporal Upscale}
\label{subsec:temporal_upscale}

In our method, the probe grid is upscaled to full resolution over multiple frames.
The amount of upscaling is directly related to the overall per-pixel sample count and can be tweaked as a trade-off between performance and quality.
In our implementation, we store up to one 8x8 screen probe, encoding the hemispherical radiance of a random pixel inside each 8x8 tile on the screen.
A consequence of this setup is that the resulting probe grid can be stored inside a 2D texture of the size of a render target, with dimensions aligned to the next multiple of 8.
This choice is arbitrary, and other configurations could be explored.

\begin{figure}
  \centering
  \begin{subfigure}{0.23\textwidth}
    \centering
    \includegraphics[width=\linewidth]{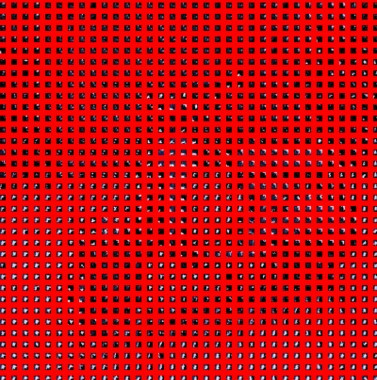}
    \caption{Sparse probe grid on 1\textsuperscript{st} frame }
  \end{subfigure}
  \begin{subfigure}{0.23\textwidth}
    \centering
    \includegraphics[width=\linewidth]{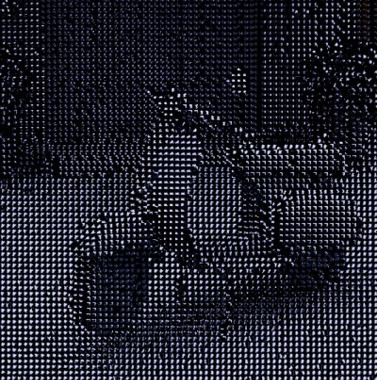}
    \caption{Resolved grid after 4 frames}
  \end{subfigure}
  \caption{Temporal upscaling in probe space. (a) One out of four probe grids are filled. Red pixels show the empty probes. In each frame, we fill one out of four. After four frames, all the probe grids are filled (b).}
  \label{fig:temporal_upscale}
\end{figure}

In this context, we define the spawn tile as a 2D region with dimensions of $8\cdot(upscale\_x,upscale\_y)$, where $upscale\_x$ and $upscale\_y$ represent the amount of temporal upscaling performed along the X and Y axis respectively.
We then generate a 2D jitter value inside the spawn tile using Halton's low-discrepancy sequence \cite{10.1145/355588.365104} and use this value to select the pixel on which to place our new probes for every spawn tile on the screen.
This leads to sparsely populated probe grids on first frames, which get resolved into fully populated grids after multiple frames. 
Figure \ref{fig:temporal_upscale} illustrates this process: on the very first frame, \textonequarter{} of all the probes are filled.
The red regions in the figure represent the tiles that do not have probe data.
We fill another \textonequarter{} in the next frame resulting in all probes being filled after four frames. 

For each probe, we reconstruct the world-space position from the depth buffer and retrieve the surface orientation by decoding the normal vector from the rasterized depth and normal textures, usually called G-buffers \cite{10.1145/97879.97901}.
Additionally, our technique requires a set of motion vectors to be calculated for every pixel to reproject the probe grid from the previous frame into the current frame.

Probe reprojection is an essential step of the pipeline, as temporally reused probes are likely to persist over multiple frames, making accurate placement of previous probes onto the new frame's pixels a requirement to prevent degrading the overall image quality.
Algorithm \ref{alg:probe_reprojection} outlines how every lane in an 8x8 dispatch group collaborates to find the best pixel for probe reuse.
An interesting input to the algorithm is the $cell\_size$ heuristic, which controls how much spatial error is allowed when reusing the probes information temporally.
As we will see in the later sections, this same heuristic is also used to guide the sampling and filtering of the screen probes, making it a key factor to get right.

\begin{algorithm}
\small
\DontPrintSemicolon
\SetKwProg{Cond}{if}{}{}
\SetKwProg{Kernel}{\_\_kernel}{($pixel$ $p$)}{}
\Kernel{reproject\_screen\_probes}{
    $\_\_local$ $uint$ $reprojection\_score \gets (pack\_half(65504.0) << 16) | 0xFFFFu$\;
    $barrier()$ \tcp{sync the threads}
    \Cond{$p$ isn't a sky pixel}{
        $q \gets p$ $in$ $previous$ $frame$\;
        \Cond{$probe_\mathrm{q}$ is valid}{
            $plane\_dist \gets abs(dot(world_\mathrm{probe} - world_\mathrm{p}, normal_\mathrm{p}))$\;
            $normal\_check \gets dot(normal_\mathrm{probe}, normal_\mathrm{p})$\;
            \Cond{$plane\_dist < cell\_size$ and $normal\_check > 0.95$}{
                $dist \gets distance(world_\mathrm{probe}, world_\mathrm{p})$\;
                $uint$ $probe\_score \gets$ \mbox{$(pack\_half(dist) << 16) | local\_lane$}\;
                $atom\_min(reprojection\_score, probe\_score)$
            }
        }
    }
    $barrier()$ \tcp{sync the threads}
    \tcp{decode and use $local\_lane$ as destination pixel}
}
\;
\caption{Screen probes reprojection.}
\label{alg:probe_reprojection}
\end{algorithm}

\subsubsection{Adaptive Sampling}

\begin{figure}
  \centering
  \begin{subfigure}{0.23\textwidth}
    \centering
    \includegraphics[width=\linewidth]{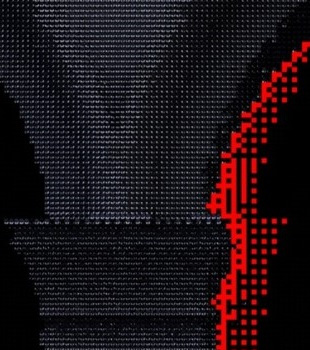}
    \caption{Fixed probe spawning pattern}
  \end{subfigure}
  \begin{subfigure}{0.23\textwidth}
    \centering
    \includegraphics[width=\linewidth]{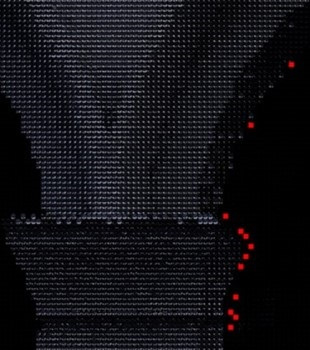}
    \caption{Stochastic ray re-balancing}
  \end{subfigure}
  \caption{Hole filling under motion with adaptive sampling.}
  \label{fig:adaptive_sampling}
\end{figure}

As a result of the sparse spawning of the screen probes, moving animations can exhibit grid regions where information is missing, as illustrated in Figure \ref{fig:adaptive_sampling}.
These regions, usually called disocclusions or holes, are detrimental to the image quality as they introduce additional noise in newly appearing parts of the scene.

One possible solution to this problem is to allocate a fixed number of additional rays to the empty tiles to "fill the holes" \cite{swoboda}.
However, such an approach requires increasing the per-pixel ray budget, which we aim to keep low and constant throughout an animated sequence.
Instead, we propose de-allocating rays randomly away from screen tiles that have succeeded the temporal reprojection and assigning them to the frame's empty tiles.

For this purpose, we generate two separate queues:
\begin{itemize}
\item The $empty\_tiles$ buffer stores the list of tiles that have failed reprojection and are not filled with any newly spawned probe.
\item The $override\_tiles$ buffer stores the list of newly spawned tiles that have succeeded the temporal reprojection.
\end{itemize}

We can patch our list of spawned tiles, referred to as $spawn\_tiles$ in Algorithm \ref{alg:adaptive_sampling}, by iterating over the empty tiles, and picking a random override tile to fill an empty tile with.
This approach enables significant improvements in the visual quality of disoccluded regions, as shown in Figure \ref{fig:adaptive_sampling}, while maintaining a constant ray count over time.
In effect, we simply redistribute a fixed ray budget to help fill our temporal holes.

\begin{algorithm}
\small
\DontPrintSemicolon
\SetKwProg{Cond}{if}{}{}
\SetKwProg{Kernel}{\_\_kernel}{($uint$ $global\_id$)}{}
\Kernel{patch\_screen\_probes}{
    $tile \gets empty\_tiles[global\_id]$\;
    $index \gets random(0, override\_tile\_count-1)$\;
    $atom\_xchg(spawn\_tiles[override\_tiles[index]], tile)$\;
}
\;
\caption{Temporal adaptive sampling.}
\label{alg:adaptive_sampling}
\end{algorithm}

\subsubsection{Ray Guiding}
\label{subsec:ray_guiding}

At this point in the pipeline, we know what probes we will be calculating.
We will be using raytracing for this but have yet to decide how to distribute our rays against the octahedral cells of each probe.
Multiple options are possible here, such as assigning one ray for every cell and randomly jittering within the cell, otherwise known as uniform sampling.

\begin{figure}
  \centering
  \begin{subfigure}{0.23\textwidth}
    \centering
    \includegraphics[width=\linewidth]{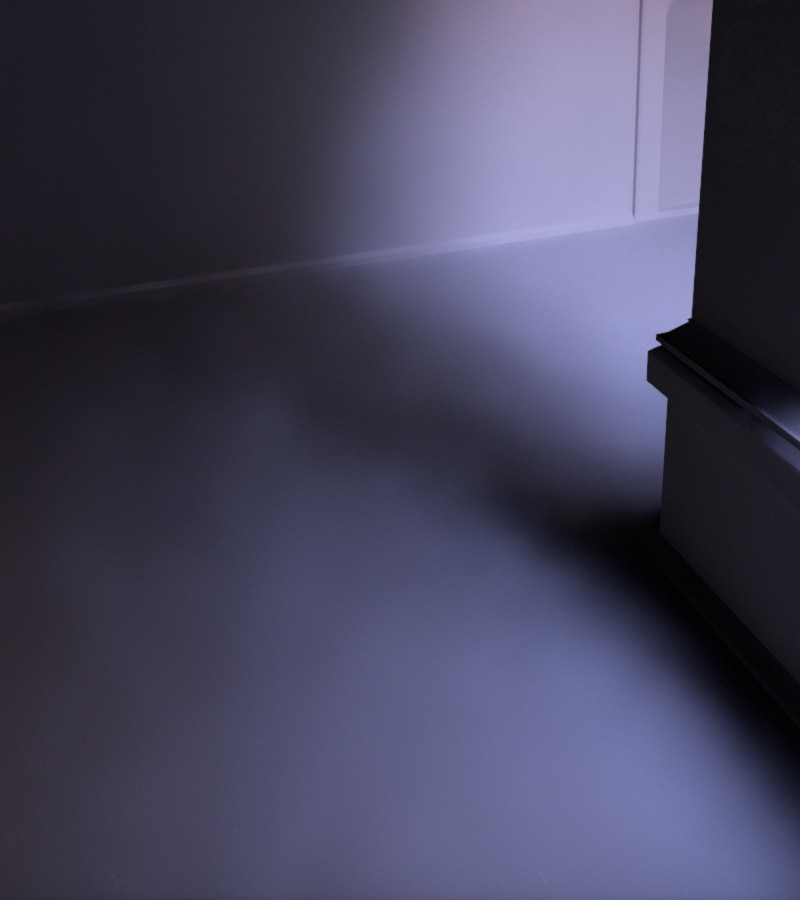}
    \caption{Uniform sampling}
  \end{subfigure}
  \begin{subfigure}{0.23\textwidth}
    \centering
    \includegraphics[width=\linewidth]{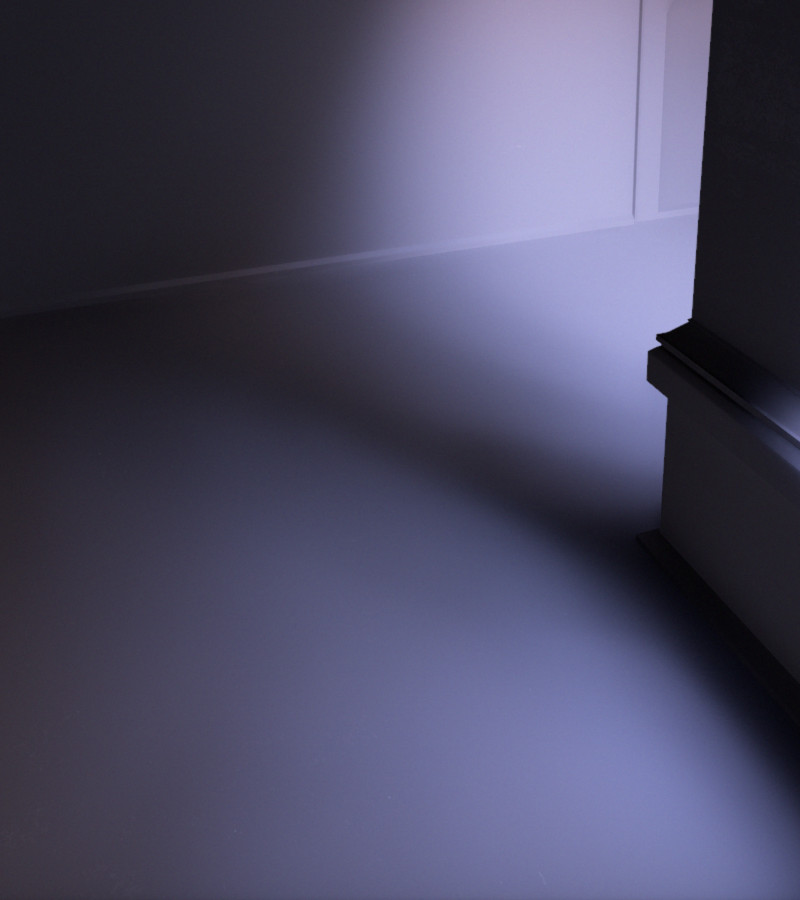}
    \caption{Temporal ray guiding}
  \end{subfigure}
  \caption{Using reconstructed hemisphere for sampling.}
  \label{fig:ray_guiding}
\end{figure}

However, we can do better than this by leveraging the information of our reprojected probe grid, when available, and guiding the sampling of the new rays or importance sampling \cite{lumen}.
We implement this efficiently by writing the luminance of the reprojected radiance to local memory, or Local Data Share (LDS); the values are then scanned in parallel \cite{harris_07} and normalized into a Cumulative Distribution Function (CDF).
We use the CDF to pick a random cell proportional to its estimated intensity and recover the ray direction by generating a random 2D sample on the selected cell.
%In particular, since the probes cover different objects with different normals and materials, we choose to shoot more rays depending on the incoming lighting.
%In fact, if we look at the rendering equation at the first intersection, we choose to do importance sampling on the lighting. 
We obtain a cleaner and more temporally stable image without increasing the ray count as illustrated in Figure \ref{fig:ray_guiding}.

\begin{figure}
  \centering
  \includegraphics[width=0.7\linewidth]{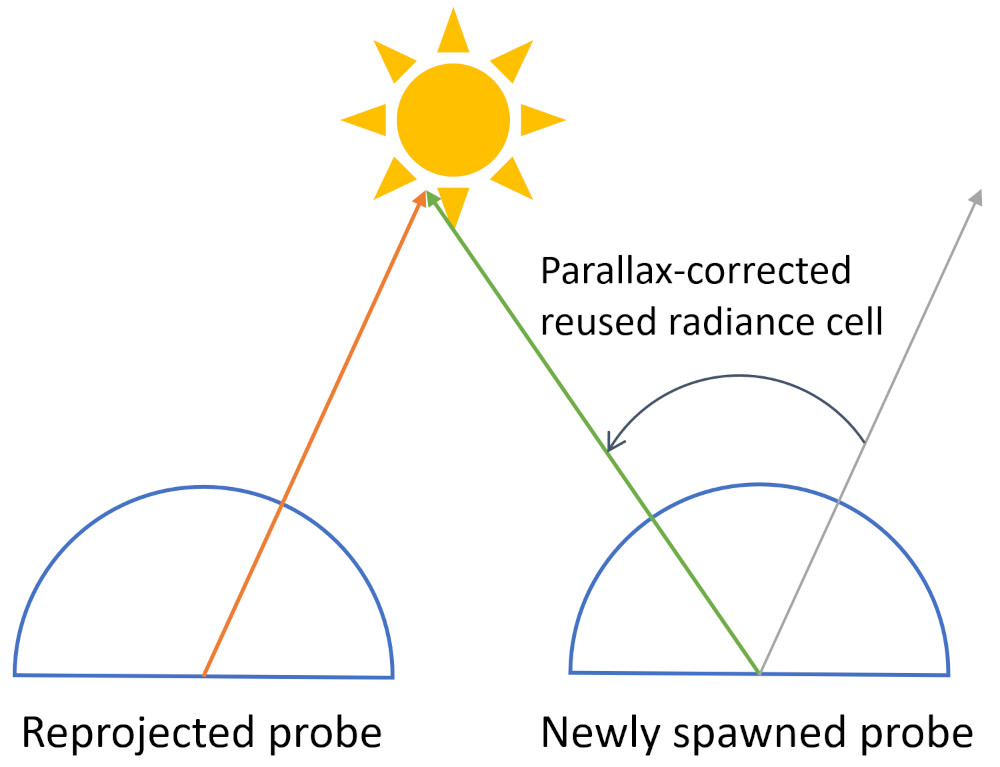}
  \caption{Parallax-corrected radiance reuse.}
  \label{fig:radiance_reuse}
\end{figure}

To make the guiding precise, we want the hemisphere reconstruction to be as faithful as possible.
Reconstruction here is iterating over the reprojected probes in a 3x3 tile neighborhood and accumulating the radiance values into the best corresponding cells of the newly spawned yet uncalculated probes.
Reusing the cell indexing across probes leads to issues with nearby light sources having a different parallax in relation to the shading site, degrading the quality of the sampling.
Therefore, we perform a parallax correction before scattering the reprojected radiance estimate into the storage of the new probe as shown in Figure \ref{fig:radiance_reuse}.
This is made possible by storing the traveled ray distance in the alpha channel of the probe grid texture, allowing recovery of the hit point position while iterating the reprojected cells.
The scattering of radiance values is implemented efficiently by allocating the 8x8 octahedral map in LDS to perform the reconstruction and sampling.
Finally, we use the same $cell\_size$ heuristic mentioned in section \ref{subsec:temporal_upscale} for rejecting far away probes.

Now that all the sampled directions are generated, rays can be intersected against the scene using a closest hit query.
If a ray misses all geometry, we consider it has reached the sky and add the corresponding incoming environment contribution to the ray payload.
If the ray hits, however, we must calculate the lighting at the hit point.
This is the role of our hash cells data structure, and we, therefore, defer the details of the implementation to section \ref{sec:hashcells}.
For now, we assume that, similarly to the environment lighting, we somehow get a radiance contribution back that we add to the ray payload.

\subsubsection{Radiance Blending}

Now that we have a radiance estimate for each of our rays, we can resolve the payload into the new probes.
As multiple rays can be assigned to the same cell, we need a way to accumulate into the destination storage efficiently.
We again leverage the LDS and allocate the 8x8 probe in local memory for accumulating the contributions prior to normalizing.

In this pass, we blend the newly calculated radiance with the estimate reconstructed in \ref{subsec:ray_guiding}.
We found that using a regular exponential moving average \cite{taa} led to a significant loss in visual fidelity.
Indeed, the low-resolution nature of the probes leads to cells covering a relatively large area of the oriented hemisphere, making averaging uniformly across the whole region undesirable.
Instead, inspired by \cite{lottes}, we adapt the temporal blending amount as a factor of the normalized difference between the newly estimated radiance and the reconstructed one.
We propose a biased temporal hysteresis detailed in Algorithm \ref{alg:temporal_hysteresis}, which we design to better preserve occlusion and shadows at the expense of some image darkening.

Further to preserving shadow details, our biased temporal hysteresis also acts as a firefly removal technique by effectively filtering out bright signals that are significantly smaller than the cone described by the cell's solid angle, making the responsible surface unlikely to be hit often.
We found this property particularly important for ensuring the temporal stability of scenes featuring many emissive surfaces or mesh lights.

\begin{algorithm}
\small
\DontPrintSemicolon
\SetKwProg{Function}{function}{}{}
\Function{temporal\_blend($curr\_radiance$, $prev\_radiance$)}{
    $l1 \gets dot(curr\_radiance, \tfrac{1.0}{3.0})$\;
    $l2 \gets dot(prev\_radiance, \tfrac{1.0}{3.0})$\;
    $\alpha \gets max(l1 - l2 - min(l1, l2), 0.0) / max(max(l1, l2), 1e-4)$\;
    $\alpha \gets clamp(alpha, 0.0, 0.95)^2$ \tcp{clamp and remap}
    \Return{$lerp(curr\_radiance, prev\_radiance, \alpha)$}
}
\;
\caption{Biased shadow-preserving temporal hysteresis.}
\label{alg:temporal_hysteresis}
\end{algorithm}

\begin{figure}
  \centering
  \begin{subfigure}{0.23\textwidth}
    \centering
    \includegraphics[width=\linewidth]{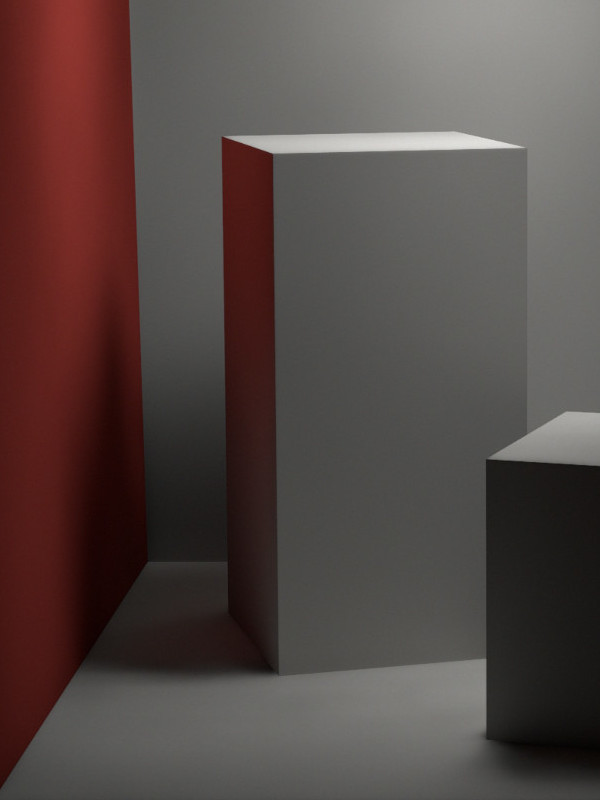}
    \caption{Darkening due to empty cells}
  \end{subfigure}
  \begin{subfigure}{0.23\textwidth}
    \centering
    \includegraphics[width=\linewidth]{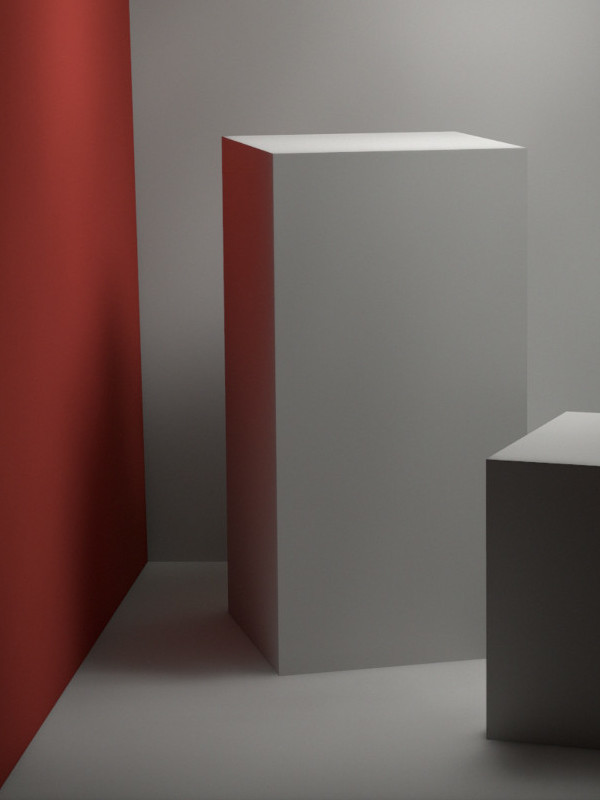}
    \caption{Fixed using radiance backup}
  \end{subfigure}
  \caption{Approximating empty cells with radiance average.}
  \label{fig:radiance_backup}
\end{figure}

Due to our ray guiding strategy, we can end up with cells for which no rays were traced.
Reusing the results of the temporal reconstruction aggressively in such cases leads to strong visual artifacts, while leaving the cell's content as black or empty leads to undesirable over-darkening caused by the missing energy.
Instead, we average the radiance from the populated cells and distribute the result uniformly across the untraced cells.
This allows recovering some of the energy loss through approximating the missing samples as illustrated in Figure \ref{fig:radiance_backup}.
In practice, the approximation is only used for low-probability cells, and we have not found this to cause objectionable visual artifacts on any of the tested content.

\subsubsection{Probe Masking}
\label{subsec:probe_masking}

\begin{figure}
  \centering
  \begin{subfigure}{0.15\textwidth}
    \centering
    \includegraphics[width=\linewidth]{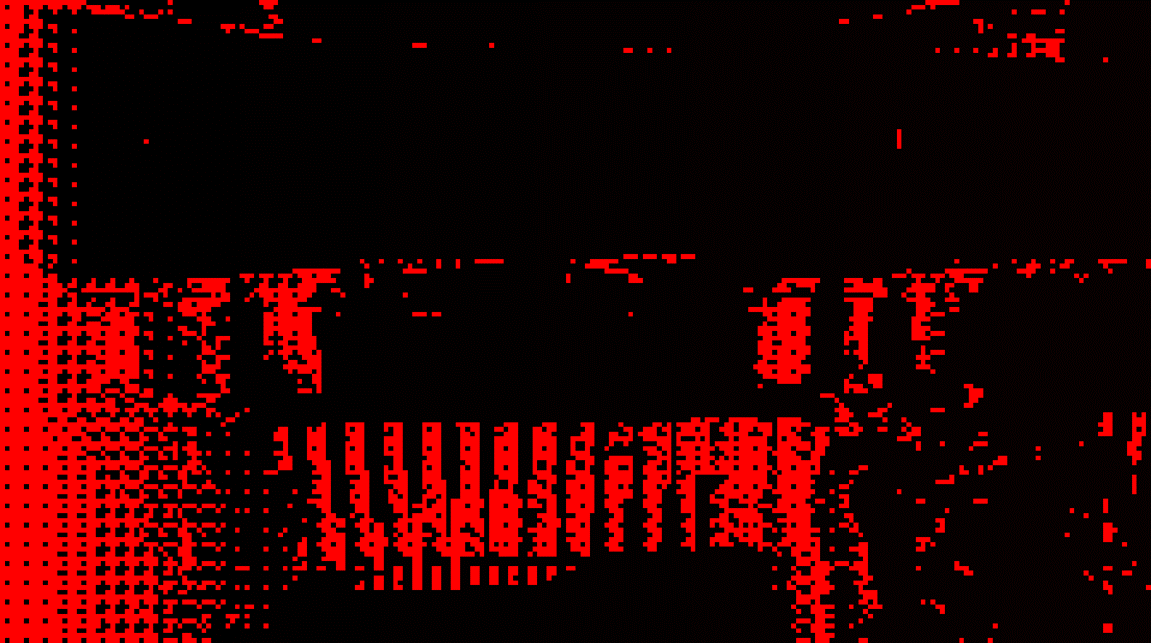}
    %\caption{...}
  \end{subfigure}
  \begin{subfigure}{0.15\textwidth}
    \centering
    \includegraphics[width=\linewidth]{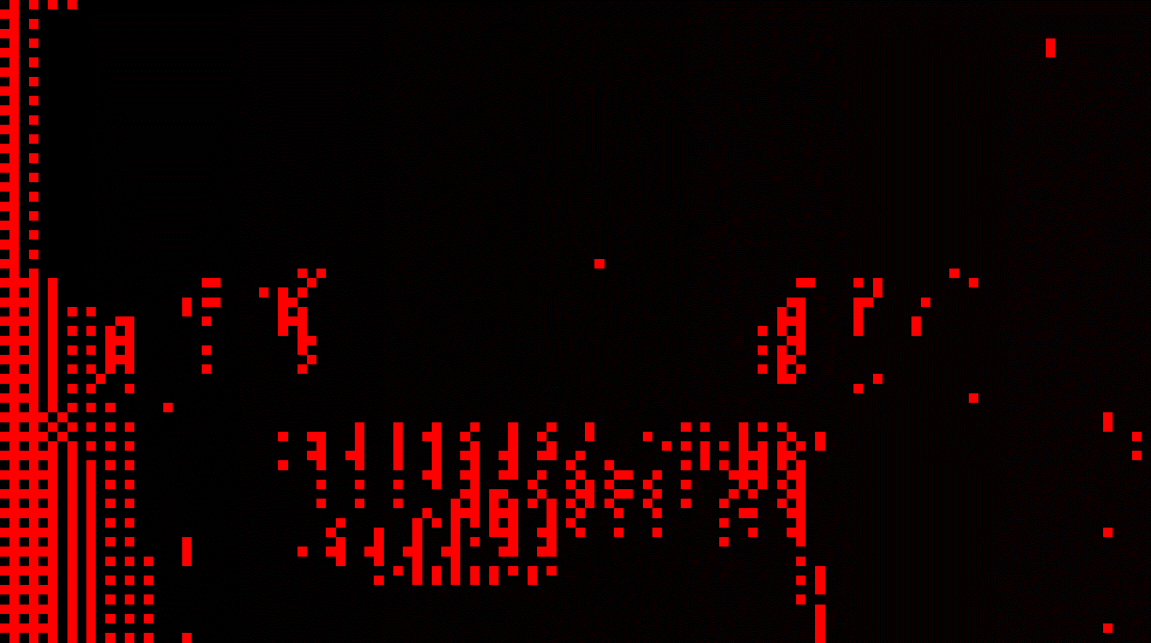}
    %\caption{...}
  \end{subfigure}
  \begin{subfigure}{0.15\textwidth}
    \centering
    \includegraphics[width=\linewidth]{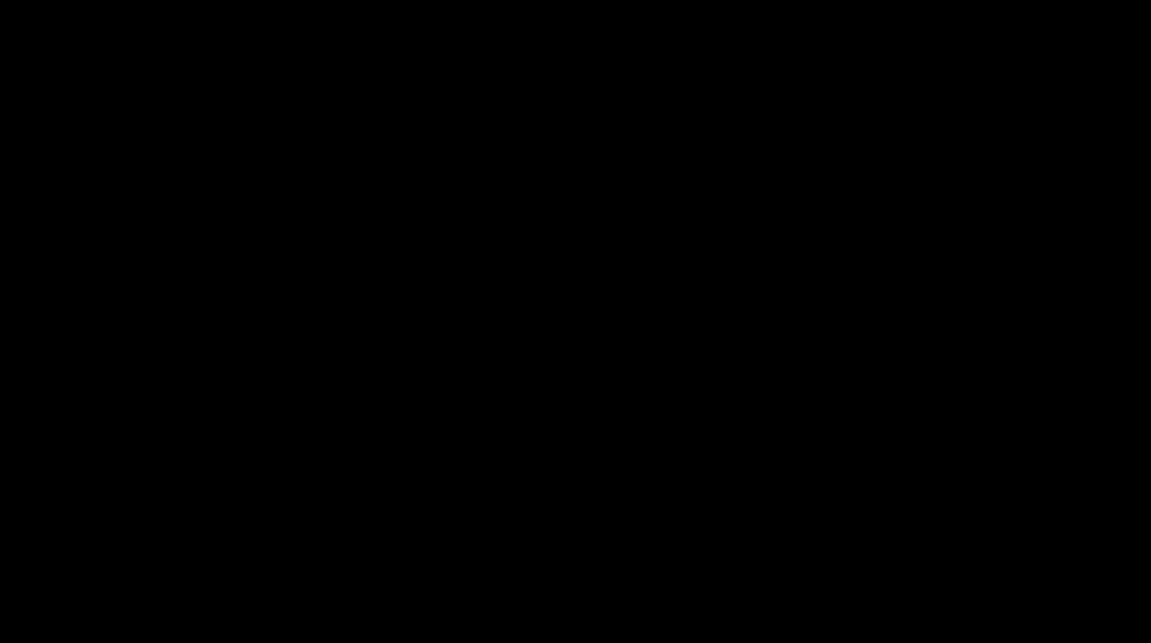}
    %\caption{...}
  \end{subfigure}
  \caption{Skipping reprojection holes w/ mip-based masking.}
  \label{fig:probe_masking}
\end{figure}

As probes can be placed on any random pixel within each of the 8x8 screen tiles, a mechanism is needed to retrieve that sub-tile position.
We therefore encode the precise pixel coordinates of each probe inside a 32-bit integer that we store in a 2D texture, referred to as $probe\_mask$.
Empty or invalid tiles, that is, tiles containing no probe, are flagged using a sentinel value, shown as red in Figure \ref{fig:probe_masking}.

As we can see in this figure, areas for which probe information is missing, can be relatively large under motion.
This is an issue as we often need to find the closest neighbor probe to a given pixel in a specific direction.
Our solution is to generate a MIP chain of the probe mask, keeping the first valid probe found within the 2x2 upper values for each successive level.
Leveraging this structure, we can efficiently perform large searches in screen space to retrieve neighbor probes in any direction, as shown in Algorithm \ref{alg:probe_search}.

\begin{algorithm}
\small
\DontPrintSemicolon
\SetKwProg{Cond}{if}{}{}
\SetKwProg{Break}{break}{}{}
\SetKwProg{Function}{function}{}{}
\SetKwProg{Foreach}{foreach}{ do}{}
\Function{find\_closest\_probe($int2$ $pixel$, $int2$ $offset$)}{
    $pixel$ $/= 8$ \tcp{transform to probe space}
    \Foreach{$mip \in 0..mip\_count-1$}{
        $int2$ $pos \gets pixel + offset$\;
        \Cond{$pos$ $is$ $out$ $of$ $bounds$}{
            \Break{}{}
        }
        $uint$ $probe \gets probe\_mask.Load(pos, mip)$\;
        \Cond{$probe$ $isn't$ $sentinel$}{
            \Return{$probe$} \tcp{found a probe :)}
        }
        $pixel$ $/= 2$\;
    }
    \Return{$sentinel$} \tcp{couldn't find any probe :(}
}
\;
\caption{Sparse directional search in probe space.}
\label{alg:probe_search}
\end{algorithm}

\subsubsection{Probe Filtering}

\begin{figure}
  \centering
  \begin{subfigure}{0.23\textwidth}
    \centering
    \includegraphics[width=\linewidth]{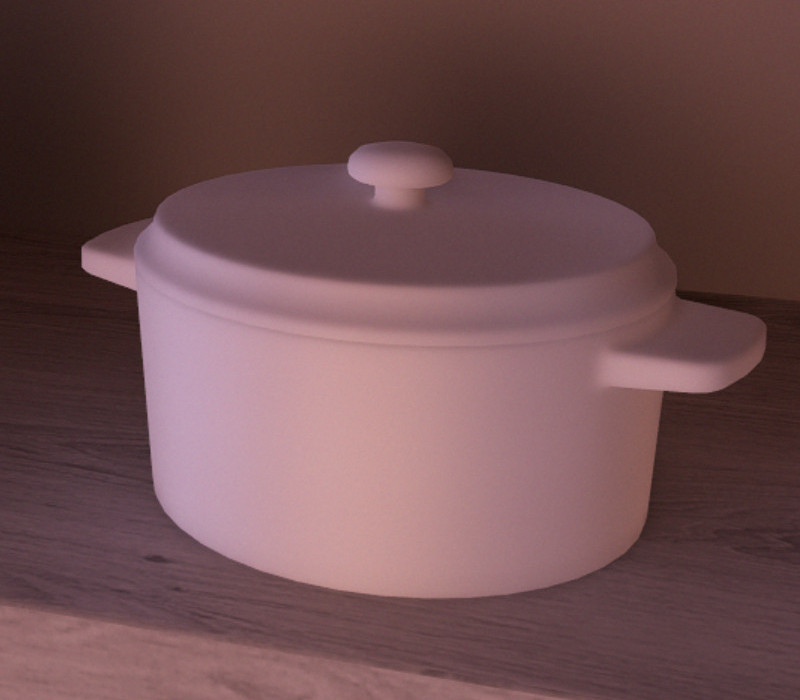}
    \caption{Naive probe-space filtering}
  \end{subfigure}
  \begin{subfigure}{0.23\textwidth}
    \centering
    \includegraphics[width=\linewidth]{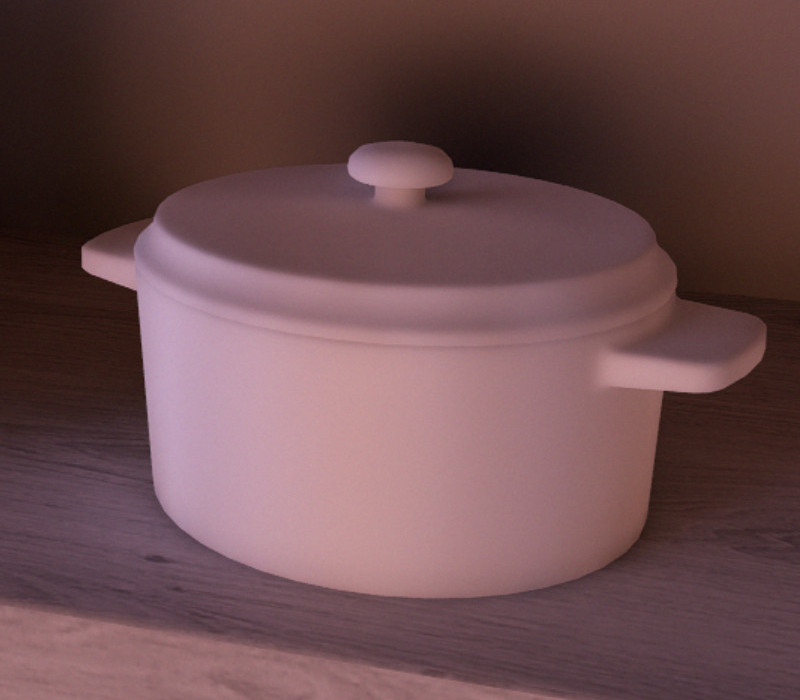}
    \caption{With angle-based rejection}
  \end{subfigure}
  \caption{7x7 separable sparse probe-space filtering.}
  \label{fig:angle_threshold}
\end{figure}

As we jitter the rays inside the cells every frame, the resulting probes can be fairly noisy.
Furthermore, the radiance returned by our hash cells may also be noisy, especially when discovering a new scene area.

We leverage our large-scale probe search function to implement an efficient separable 7x7 sparse blur of the probes' radiance, as shown in Algorithm \ref{alg:probe_filtering}.
The $cell\_size$ heuristic is again used to avoid filtering across far away probes and prevent light leaking.
Finally, we use a similar angle error detection to \cite{lumen} for preserving small-scale occlusion details as illustrated in Figure \ref{fig:angle_threshold}.

\begin{algorithm}
\small
\DontPrintSemicolon
\SetKwProg{Cond}{if}{}{}
\SetKwProg{Continue}{continue}{}{}
\SetKwProg{Foreach}{foreach}{ do}{}
\SetKwProg{Kernel}{\_\_kernel}{($global\_id$, $local\_id$, $group\_id$)}{}
\Kernel{filter\_screen\_probes}{
    $p \gets decode\_probe\_mask(spawn\_tiles[group\_id])$\;
    $dir \gets calculate\_cell\_direction(local\_id, normal_\mathrm{p})$\;
    $radiance \gets probe\_buffer[global\_id]$\;
    $hit\_dist \gets radiance.a$\;
    $weight \gets 1.0$\;
    \Foreach{$i \in 0..5$}{
        $step \gets (((i$ $\&$ $1) << 1) - 1) \cdot ((i >> 1) + 1)$\;
        $probe \gets find\_closest\_probe(p, step \cdot blur\_direction)$\;
        \Cond{$probe$ $is$ $sentinel$}{
            \Continue{}{}
        }
        $q \gets decode\_probe\_mask(probe)$\;
        $plane\_dist \gets abs(dot(world_\mathrm{q} - world_\mathrm{p}, normal_\mathrm{p}))$\;
        $normal\_check \gets dot(direction, normal_\mathrm{q})$\;
        \Cond{$plane\_dist > cell\_size$ or $normal\_check < 0.0$}{
            \Continue{}{}
        }
        $hit\_dist\_clamped \gets min(hit\_dist_\mathrm{q}, hit\_dist)$\;
        $hit\_point \gets world_\mathrm{q} + dir \cdot hit\_dist\_clamped$\;
        $angle\_error \gets dot(dir, normalize(hit\_point - world_\mathrm{p}))$\;
        \Cond{$angle\_error < cos(\tfrac{\pi}{50.0})$}{
            \Continue{}{}
        }
        $depth\_weight \gets calculate\_depth\_weight(depth_\mathrm{p}, depth_\mathrm{q})$\;
        $radiance$ $+= depth\_weight \cdot float4(radiance_\mathrm{q}, hit\_dist\_clamped)$\;
        $weight$ $+= depth\_weight$\;
        $hit\_dist \gets \tfrac{radiance.a}{weight}$\;
    }
    \tcp{store $\tfrac{radiance}{weight}$}
}
\;
\caption{Radiance filtering in probe space.}
\label{alg:probe_filtering}
\end{algorithm}

\subsubsection{Adaptive Cell Size}
\label{subsec:adaptive_cell_size}

\begin{figure}
  \centering
  \begin{subfigure}{0.23\textwidth}
    \centering
    \includegraphics[width=\linewidth]{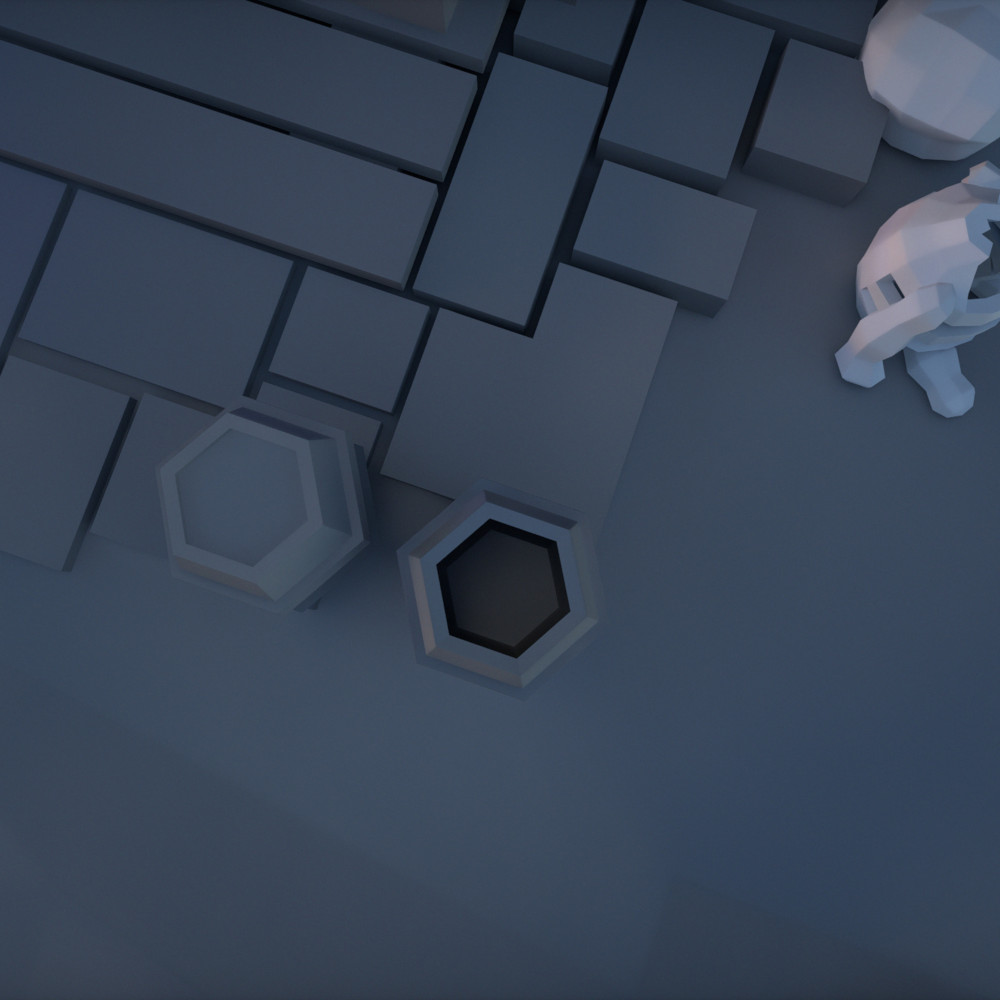}
    \caption{Details are captured up-close}
  \end{subfigure}
  \begin{subfigure}{0.23\textwidth}
    \centering
    \includegraphics[width=\linewidth]{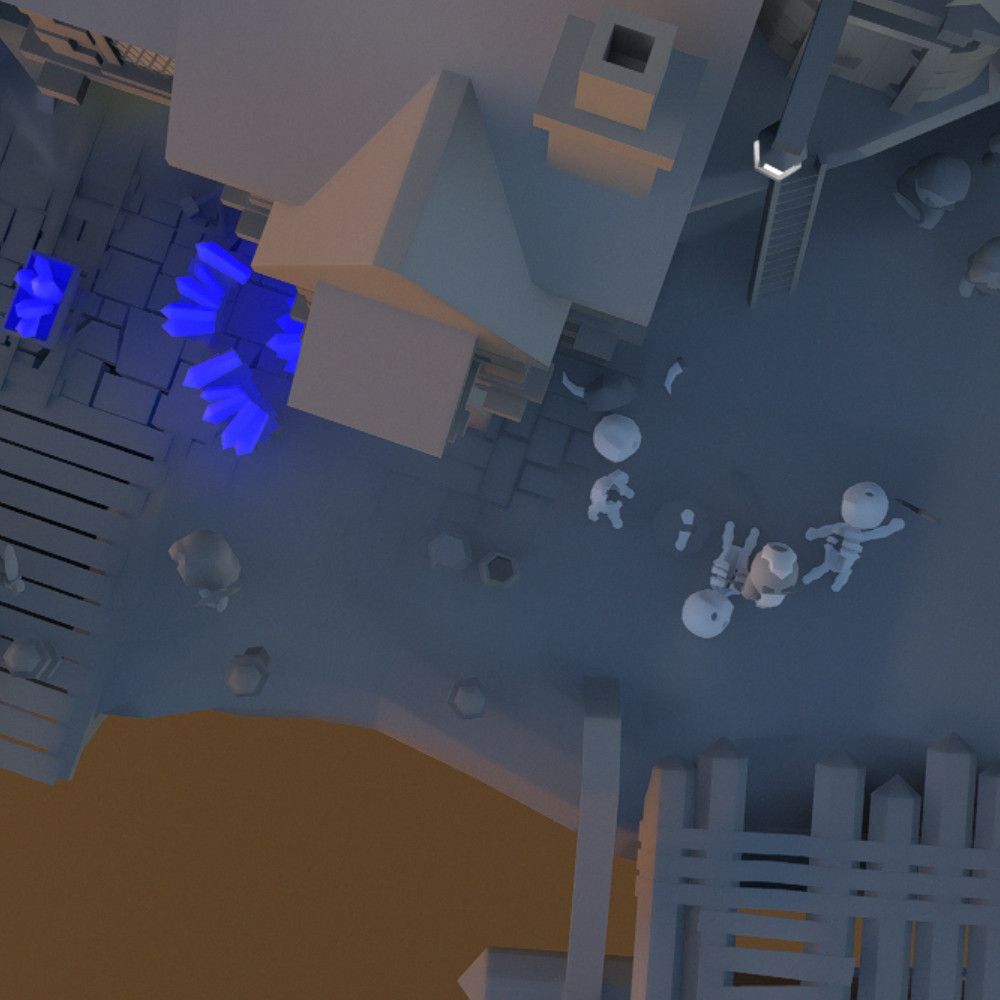}
    \caption{Gracefully degraded far out}
  \end{subfigure}
  \caption{Relaxing radiance reuse heuristics at a distance.}
  \label{fig:adaptive_cell_size}
\end{figure}

Throughout the previous sections, we have used the same heuristic referred to as $cell\_size$ for guiding the reprojection, sampling, and filtering of our screen probes.
This quantity directly relates to how permissive the radiance reuse is between neighboring probes; a low value leads to better detail preservation at the cost of degraded temporal stability, while a high value helps maintain temporal stability but at the cost of decreased lighting quality.
Therefore, we choose to use an adaptive value relaxed for objects located further away from the camera, as illustrated in Figure \ref{fig:adaptive_cell_size}.

Algorithm \ref{alg:adaptive_cell_size} describes the calculations for the adaptive cell size, where $fov\_y$ corresponds to the vertical field of view in radians, and $proj\_size$ is the targeted cell size in pixels after projection.
We choose $proj\_size$ to be set to a value of $8.0$, which is roughly the amount of pixel spacing between neighbor probes.
Note that $distance\_scale$ may be precalculated on the CPU as it is constant across the screen.

\begin{algorithm}
\small
\DontPrintSemicolon
\SetKwProg{Function}{function}{}{}
\Function{calculate\_cell\_size($distance\_to\_camera$)}{
    $distance\_scale \gets tan(fov\_y \cdot proj\_size \cdot max(\frac{1.0}{view\_height}, \frac{view\_height}{view\_width^2}))$\;
    \Return{$distance\_scale \cdot distance\_to\_camera$}
}
\;
\caption{Calculating an adaptive cell size.}
\label{alg:adaptive_cell_size}
\end{algorithm}

\subsubsection{Persistent Least-Recently Used (LRU) Side Cache}

Thanks to the temporal upscale, our screen cache is relatively dense across the screen over throughout multiple frames.
However, despite many probes, we are still heavily undersampled with only one screen probe available for each 8x8 pixel tile.
This is not an issue in itself as probes are meant to be interpolated from in a process where each pixel locates its four closest neighbors and performs some form of the average of the irradiance signal.
We will cover the details of our probe interpolation and irradiance estimation in section \ref{subsec:irradiance_estimation}.

This section focuses on the temporal instability issues introduced by thin geometrical features.
We have mentioned that the $cell\_size$ heuristic is used to choose whether to keep or reject temporal probe history and how we adapt this value at a distance to account for far away regions having a lower ratio of sample count to geometrical detail.
However, there exist situations where the spatiotemporal radiance reuse between probes keeps failing, causing temporal wobbling in the illumination.

\begin{figure}
  \centering
  \includegraphics[width=0.5\linewidth]{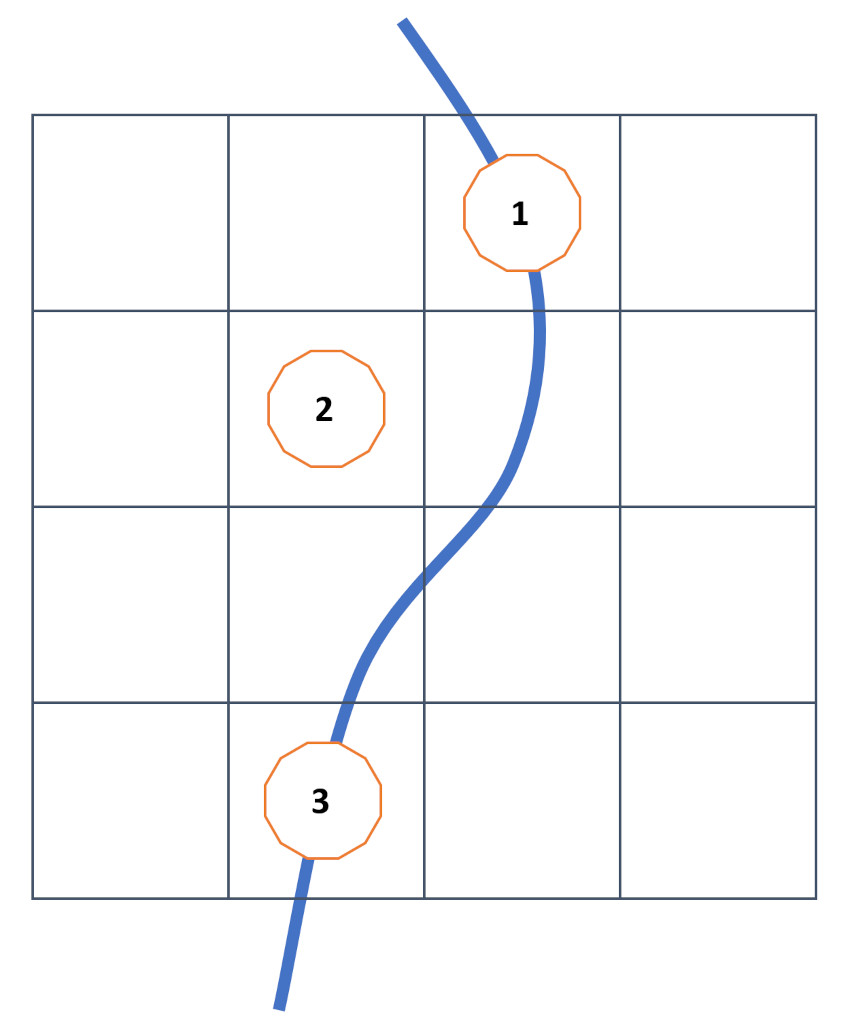}
  \caption{Temporal reuse failure caused by thin geometry.}
  \label{fig:lru_side_cache}
\end{figure}

Figure \ref{fig:lru_side_cache} describes such a scenario:
\begin{itemize}
    \item Probe $1$ is initially spawned on the thin geometrical feature and calculated.
    \item Probe $2$ is then spawned but fails to reuse the temporal information from $1$ as the $cell\_size$ test fails.
    It is therefore recalculated from scratch or, in other words, no ray guiding nor radiance blending can be performed.
    It is worth noting that, as mentioned in section \ref{subsec:ray_guiding}, we perform the reconstruction against a 3x3 tile neighborhood in probe space rather than a single tile.
    However, we will ignore this property here to simplify the explanations.
    \item Finally, probe $3$ is spawned but cannot reuse information from $2$ as it is back on the geometrical feature.
\end{itemize}

The net result of this highlighted scenario is that our temporal reprojection, ray guiding, and radiance blending keep failing each frame, leading to an unstable lighting response.
However, we can note that we had valid temporal information for reconstructing probe $3$ just two frames ago.
We, therefore, propose to push the "evicted probes" onto a persistent queue so that such probes may be reused an arbitrary number of frames later.

In this context, evicted probes are detected when newly spawned probes fail to reuse data from the reprojected probe within the same tile during reconstruction.
The queue itself is implemented as a simple array of probe indices, while the radiance data is stored in another 2D texture with identical dimensions to our regular probe grid.
This choice is arbitrary, and other configurations could be explored.

In order to use what we will call the "cached probes" as part of the reconstruction, we need a mechanism for efficiently looking up the nearby probes for a given pixel.
We implement a lookup structure similar to that used in tile-based deferred rendering pipelines \cite{tilebaseddeferred} by iterating over the list of cached probes prior to the reconstruction, projecting them onto the screen, and scattering the probe indices into the dedicated list of the corresponding 8x8 tile.
Since cached probes can persist over many frames, we cannot rely on the depth buffer to recover the world-space position. Instead, we store the single-precision floating-point position and packed world-space normal into a 128-bit integer for each cache entry.

\begin{figure*}
  \centering
  \includegraphics[width=\linewidth]{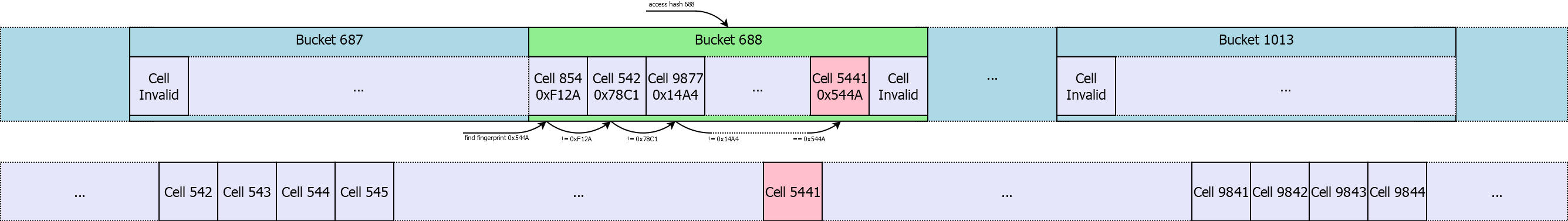}
  \caption{Hash map with open addressing and linear probing.
  %Buckets form a finite array that are indexed using a hash function. Cells form another array indexed by the bucket entries. Bucket entries also store cell fingerprints for linear probing.
  }
  \label{fig:hash}
\end{figure*}

We update our hemisphere reconstruction routine to search for cached probes in a 3x3 tile neighborhood after completing the 3x3 tile search on the reprojected grid.
All radiance contributions are still parallax-corrected and accumulated into LDS as before.
Once the search is completed, we end up with three possible scenarios with regard to updating our side cache:
\begin{itemize}
    \item The previous probe is evicted by the new one, and no matching cache entry was found; we create a new cached probe and schedule the reprojected radiance to be written out to the cache during the blending pass.
    \item The previous probe is evicted by the new one but a matching cache entry could be found; we schedule that the reprojected radiance be written out to the cache during the blending pass and push the entry onto a most recently used (MRU) queue.
    \item Regardless of whether the previous probe was evicted, we identify the best-matching cache entry that participated in the reconstruction, if any, and schedule the new radiance estimate to be written out to the cache during the blending pass.
    This step is essential, so the cached radiance does not lag under changing lighting conditions.
\end{itemize}

We mentioned an MRU queue in the case of finding a match for an evicted probe.
We indeed perform a reordering pass of the indices of the cached probes every frame, keeping the MRU elements ahead while the LRU entries naturally fall behind.
This setup allows to avoid evicting cache entries that would be actively participating in illuminating the current viewpoint.

Finally, it is worth mentioning that we perform an atomic compare and swap operation before scheduling of the cache updates mentioned in the above scenarios.
This is to avoid a race condition where multiple work groups may try and update the same cached content during the radiance blending.
Additionally, care must be taken as entries scheduled for update may have been overwritten by newly created cache entries. This can be verified trivially by checking the entry's index inside the LRU queue against the allocation cursor for the newly created cache entries.

\subsection{World Cache}
\label{sec:hashcells}

We previously skipped over the details of calculating and caching the outgoing radiance at each of the secondary path vertices.
This is the role of our world cache, stored in hash cells, where we accumulate and filter the radiance being returned to the screen cache queries.
We describe in this section how we build on the work of \cite{cite:mppsf}, using spatial hashing to create our data structure, which we further extend to allow fast filtering across neighbor cells using a novel tiling approach.

\subsubsection{Caching Outgoing Radiance for Secondary Path Vertices}
\label{subsec:radiance_caching}

We aim at minimizing the memory footprint while requiring little to no preprocessing of the scene geometry.
The latter is an important requirement for supporting dynamic worlds as well as facilitating our solution's integration into an existing real-time rendering pipeline.
\cite{cite:mppsf} proposes to cache the radiance into cells using spatial hashing.
Their approach provides interesting properties that we want to build on:
\begin{itemize}
    \item Their use of spatial hashing adapts to any geometrical input and moving content.
    \item Filtering can be performed at no additional cost inside each cell thanks to the scattering logic used when writing out the radiance samples.
    \item The resulting grid is built entirely on the fly, as traversal occurs, therefore only requiring to allocate memory sparsely where information is needed.
\end{itemize}

Our world cache addresses the radiance cells by hashing a descriptor for each vertex.
The descriptor, which we will describe next, is hashed using a first hash function to retrieve the index of a "bucket" as shown in Figure \ref{fig:hash}.
Then, it is hashed a second time using another hash function to calculate a "fingerprint" which we use to perform linear probing inside the bucket and locate our cell in memory.
Linear probing is an essential step in solving the collisions that arise when two distinct descriptors wrongly resolve to the same bucket after only applying the initial hashing.
Finally, we leverage the study by \cite{Jarzynski2020Hash} and pick two fast hash functions that we found to produce little to no collision between one another.

\begin{figure*}
  \centering
  \includegraphics[width=\linewidth]{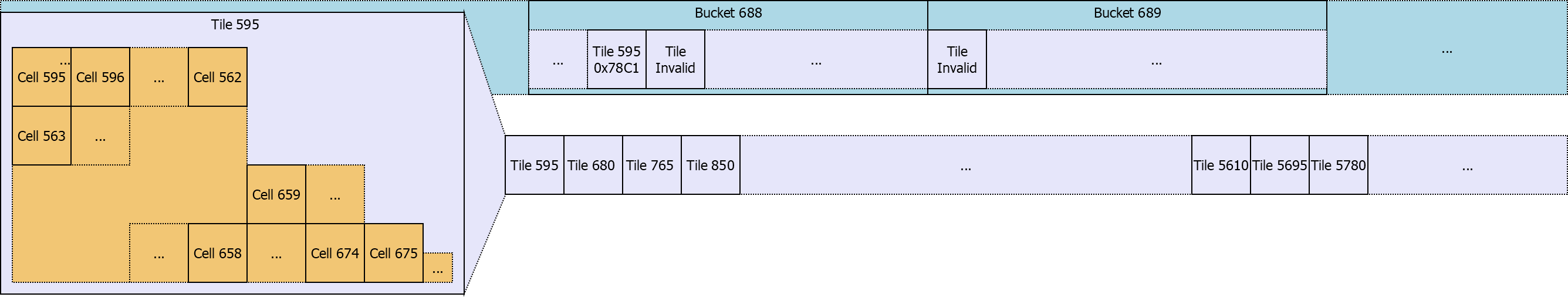}
  \caption{Two-level hash map. Each tile is a mipmapped 8x8 region of cells. Tiles and cells follow a linear layout in memory.}
  \label{fig:hash_2_levels}
\end{figure*}

As we want to cache the outgoing radiance inside the cells of our structure, we define our descriptors as the world space position of each vertex along with the ray's direction or, in other words, the input parameters to the rendering equation \cite{10.1145/15886.15902}.
In practice, both attributes are quantized in order to enable reuse and filtering across neighboring vertices.
Additionally, similarly to the adaptive size described in section \ref{subsec:adaptive_cell_size}, we adapt the amount of quantization applied to the position attribute at a distance to ensure a roughly constant number of samples per cell.
The descriptor is updated to include the quantization level, effectively creating a radiance level of details (LODs).

Each cell is associated with a decay value, which is reset upon access, or left to decay toward zero otherwise.
As a cell's decay reaches zero, we deallocate the entry so its memory can be later reused for another region of world space.
The main bottleneck of a global hash map, such as the one we are describing, is the extensive use of atomic operations for each insertion into the cache.
While these operations remain relatively fast, we want to minimize their use for efficiency reasons.
In practice, we only perform a single lookup for each vertex and cache the resulting index out to memory.

\subsubsection{Eliminating Light Leaks}

\begin{figure}
  \centering
  \includegraphics[width=\linewidth]{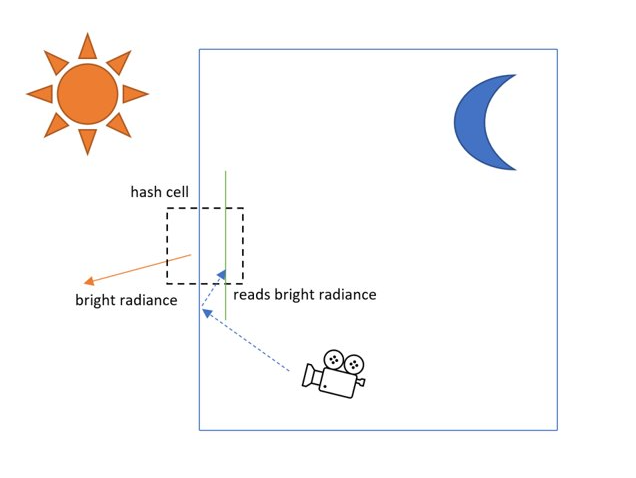}
  \caption{
  Illustrating a possible light leaking scenario.
  %The cell is split by a wall between a dark room and a bright outdoor. Both outgoing direction and positions are close and if we don't split the cell using the boolean heuristic some bright radiance from the outdoor will leak into the room.
  }
  \label{fig:light_leak_figure}
\end{figure}

\begin{figure}
  \centering
  \begin{subfigure}{0.23\textwidth}
    \centering
    \includegraphics[width=\linewidth]{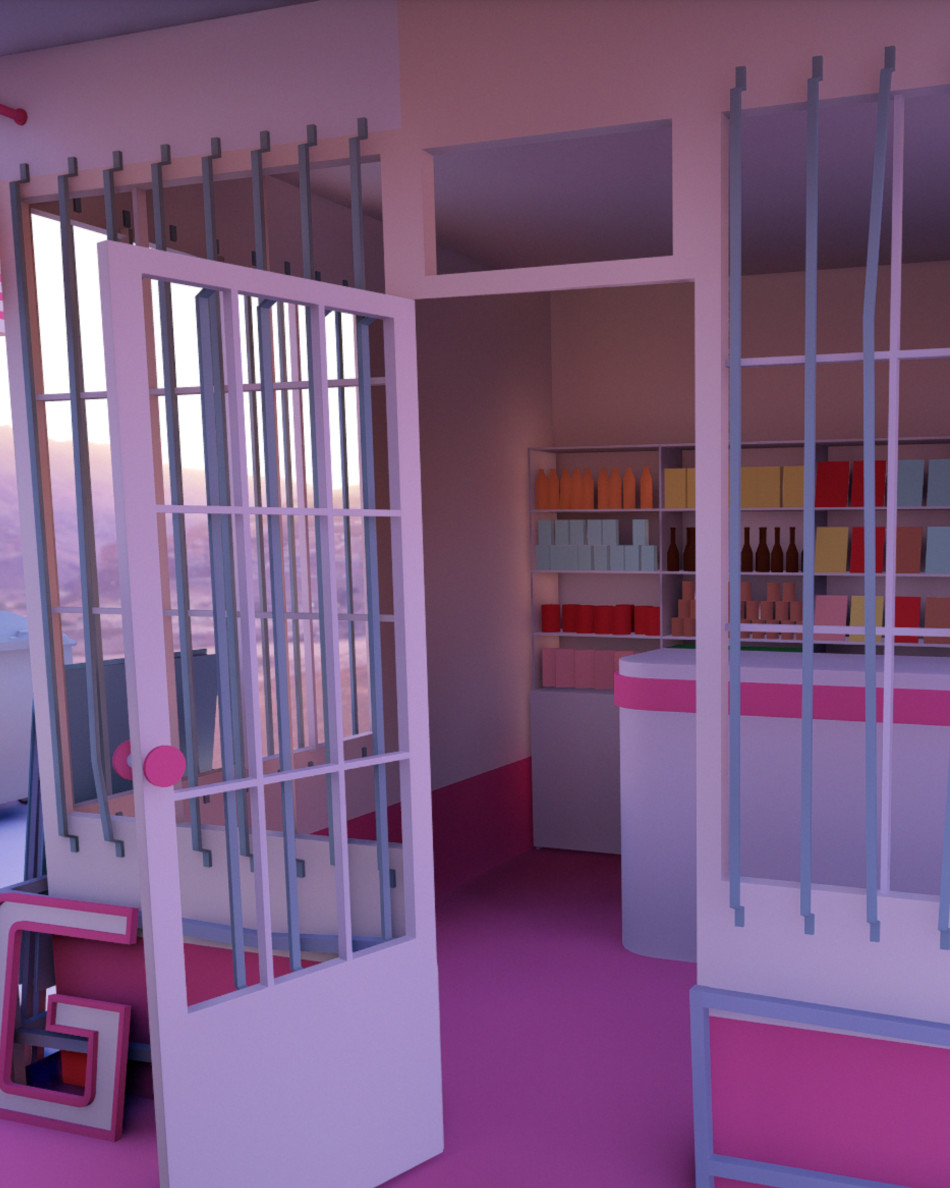}
    \caption{Without light leak fixup}
  \end{subfigure}
  \begin{subfigure}{0.23\textwidth}
    \centering
    \includegraphics[width=\linewidth]{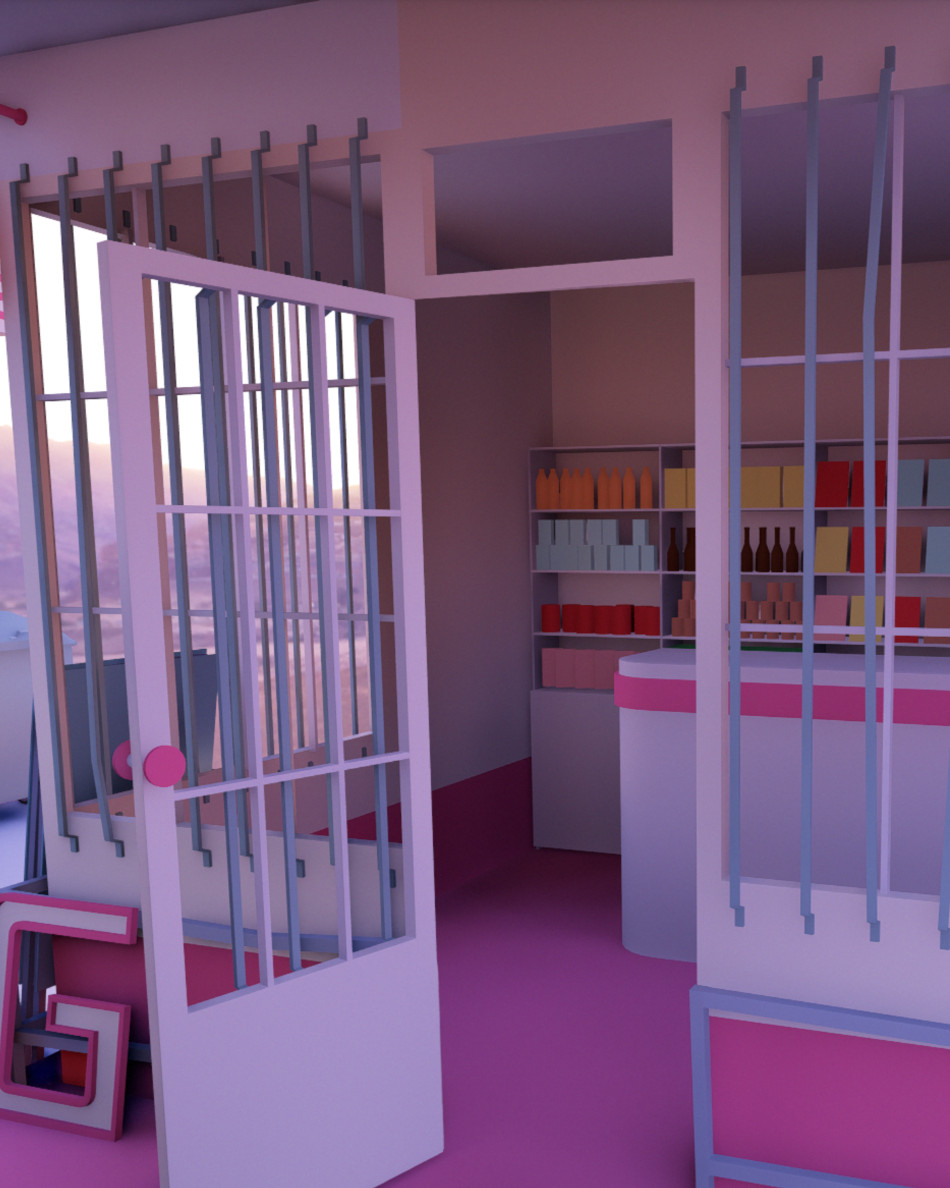}
    \caption{With light leak heuristic}
  \end{subfigure}
  \caption{Fixing light leaks caused by thin occluders.}
  \label{fig:light_leaks}
\end{figure}

The quantization of the descriptor attributes mentioned in the previous section allows for efficient filtering of the radiance signal across neighbor vertices, while the adaptivity of the quantization amount ensures a roughly constant number of samples per cell at any distance.
However, this same quantization may be responsible for light-leaking artifacts under certain conditions.
Figure \ref{fig:light_leak_figure} illustrates such a scenario.
Here, the inner "green wall" is reached by our secondary ray after bouncing off the outer wall.
The hit point belongs to an enclosed dark region of the scene, and the estimated radiance should be dark too.
However, we also see a "bright radiance" event happening on the wall's outer side; here, the environment is bright and sunny.
In this setup, both the bright and dark radiance events are close enough in world space to equal their positions after quantization.
However, we have seen that our descriptors also include the ray direction, which can help break apart two nearby vertices into different cells.
In this scenario, though, both the secondary ray and bright radiance direction are parallel enough that their directions are equal after quantization.
Our secondary ray hit, therefore, shares the same cell after hashing as the bright radiance event, resulting in a light leak as illustrated in Figure \ref{fig:light_leaks}.

We found that such light-leaking situations mainly occur when the length of the secondary ray is less than the size of the cell being looked up, or, in other words, the ray hasn't traveled for more than a cell.
We, therefore, enhance our descriptor by additionally hashing the boolean result of the $(ray\_length < cell\_size)$ inequality test.
This helps break the two events mentioned above into two distinct cells fixing our leaking issue as seen in Figure \ref{fig:light_leaks}.

\subsubsection{Prefiltering Radiance}
\label{subsec:prefiltering_the_radiance}

Filtering across cells requires quickly accessing neighbors; performing additional lookups in even just a 3x3x3 hash-space neighborhood ends up being prohibitively expensive, so we propose implementing a two-level data structure.
In this context, cells are not directly indexed by the buckets but rather grouped into tiles, as illustrated in Figure \ref{fig:hash_2_levels}.
Indexing a cell requires its tile index and relative position inside the tile.
Each tile contains a fixed number of cells that we can store in memory using a linear layout.
This layout also includes MIP levels, which we use for prefiltering the radiance, thus enabling fast spatial filtering across cells within a tile.
Tiling doesn't help, however, in filtering the radiance between tiles, which remains expensive. Instead, we rely on our screen cache to hide the resulting structured artifacts at no additional cost.

Each tile represents a fixed volume of the scene; however, a 3D encoding of the cells isn't necessary for caching and filtering the radiance.
Indeed, surfaces tend to form a plane when seen at a small enough scale.
Thus, a 2D tile of cells is sufficient for faithfully describing the radiance of a local region of space.
We build our tiles by projecting along the main axis for indexing the cell within its tile.
The largest component of the outgoing direction defines the main axis, and the projection is made by discarding the corresponding coordinate when computing the cell offset within its parent tile.
This 2D rather than 3D tiling approach helps maximize the tile occupancy, or, in other words, the ratio of used versus unused cells. This ultimately helps reduce the overall memory budget and keep the filtering fast.

Reducing the variance to an acceptable level typically requires several frames worth of contributions inside a cell.
The outgoing radiance, therefore, needs to be filtered before returning to the screen cache queries, especially for cells that have been recently created.
Contributions are temporally accumulated into the cells at the first MIP level using an exponential moving average \cite{taa}.
Other MIP levels are then iteratively generated using box filtering of radiance samples from the upper level in a single pass.
In practice, we found that the best performance was achieved with a tile size of 8x8.
Finally, we accumulate the radiance into the screen cache queries using the first finest level for which enough samples have been accumulated. 

\subsubsection{Evaluating Lighting at Secondary Path Vertices}
\label{subsec:radiance_feedback}

\begin{figure}
  \centering
  \begin{subfigure}{0.23\textwidth}
    \centering
    \includegraphics[width=\linewidth]{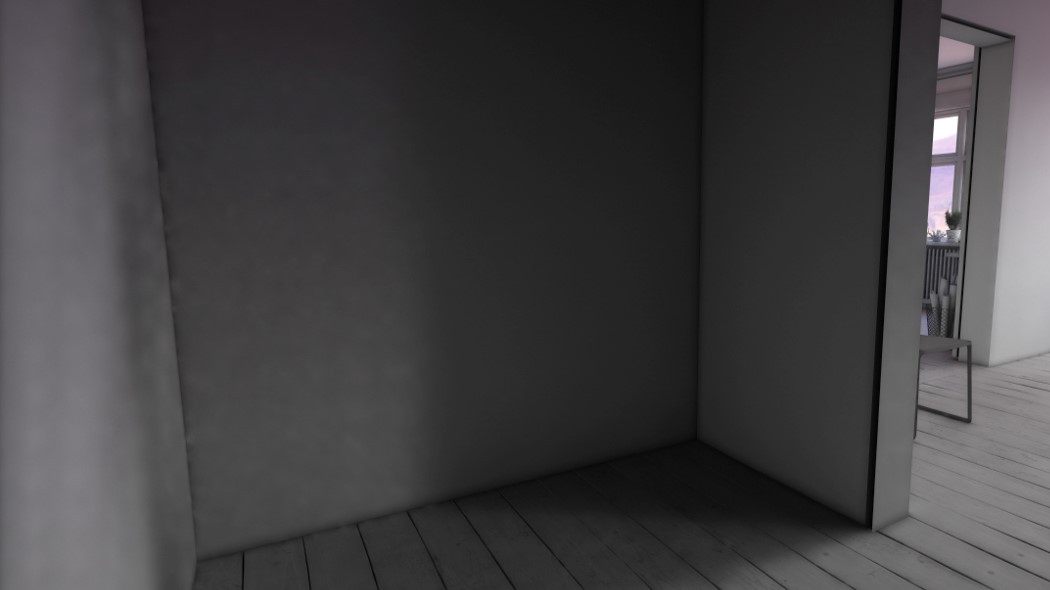}
    \caption{Using only shadow rays}
  \end{subfigure}
  \begin{subfigure}{0.23\textwidth}
    \centering
    \includegraphics[width=\linewidth]{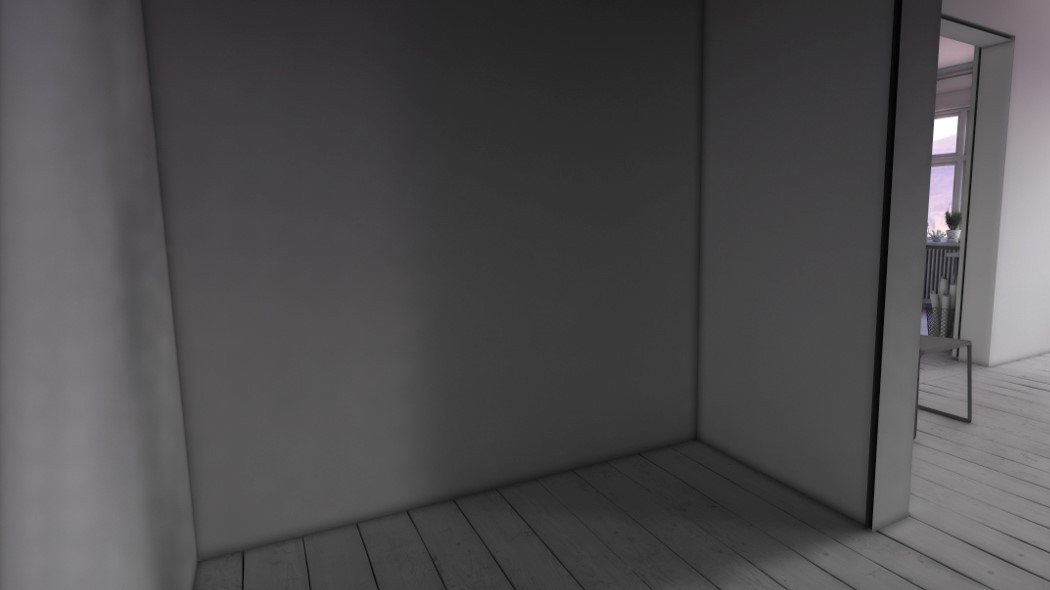}
    \caption{With radiance feedback}
  \end{subfigure}
  \caption{Approximate multi bounce w/ temporal feedback.}
  \label{fig:radiance_feedback}
\end{figure}

We have described how we cached and filtered the direct lighting at secondary path vertices inside a two-level hash grid data structure.
However, the direct lighting still needs to be evaluated.
First, we reuse the illumination from the last frame using temporal reprojection.
If the reprojection succeeds, we can reuse the radiance sample and forego any further, potentially expensive calculations.
However, if the reprojection fails, we need a reliable way of estimating the lighting at the vertex site. This is the role of our light sampling implementation, which we describe in section \ref{subsec:light_sampling}.
It is worth noting that when reprojecting the last frame's lighting, the reused radiance sample encodes the direct lighting as well as an indirect lighting estimate.
This is an interesting property, which leads to what we refer to as "temporal radiance feedback" and allows us to approximate an infinite number of light bounces as seen in Figure \ref{fig:radiance_feedback}.
This approximative multi-bounce illumination is estimated at no additional cost over multiple frames.
Finally, it is worth noting that the result of the reprojection is filtered, thus injecting a noise-free radiance sample into our world cache, leading to an interesting reduction in the variance of the resulting indirect illumination.

\subsection{Light Sampling}
\label{subsec:light_sampling}

\begin{figure}
  \centering
  \begin{subfigure}{0.23\textwidth}
    \centering
    \includegraphics[width=\linewidth]{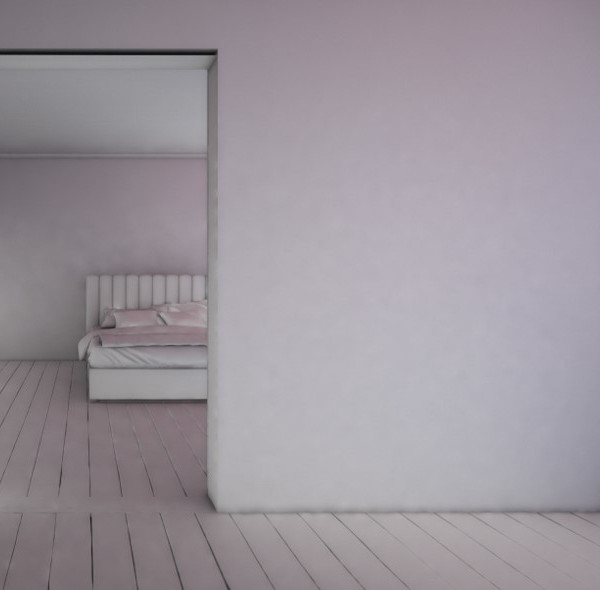}
    \caption{Blotches from hash cells noise}
  \end{subfigure}
  \begin{subfigure}{0.23\textwidth}
    \centering
    \includegraphics[width=\linewidth]{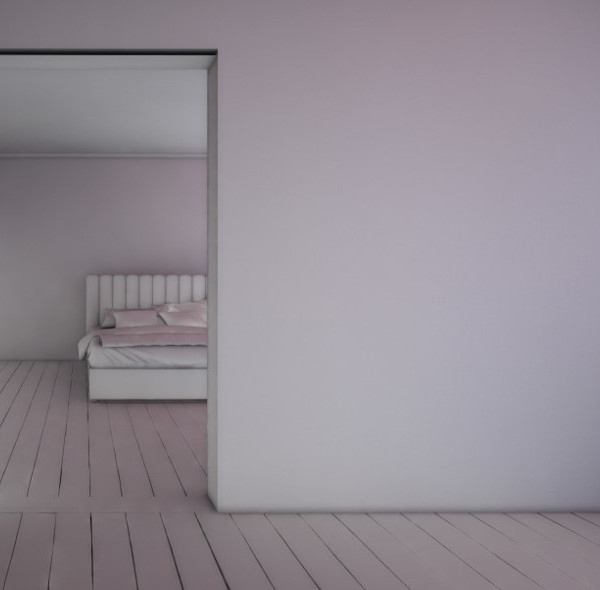}
    \caption{Fixed w/ world-space ReSTIR}
  \end{subfigure}
  \caption{Using ReSTIR for efficient light sampling.}
  \label{fig:restir}
\end{figure}

We have seen that populating the radiance cache requires evaluating the direct lighting at each of our hit points.
Our implementation supports traditional real-time visibility techniques such as shadow maps; however, such techniques typically do not scale to large numbers of lights.
We, therefore, propose a light sampling strategy, using raytracing, to stochastically sample the lighting distribution and fill the radiance cache.

\subsubsection{Reservoir-based Resampling}
\label{subsec:restir}

Sampling the direct lighting distribution can lead to large variance when a significant proportion of sampled shadow rays do not hit the light source, therefore, failing to contribute meaningfully to the estimate.
This can strongly impact the world cache variance which flows through to the screen cache and ultimately results in increased onscreen noise as illustrated in Figure \ref{fig:restir}.
Therefore, efficient importance sampling of the lighting distribution is key to keeping the noise to a minimum while casting as few rays as possible.

%This requires a sampling strategy that can maintain the lowest possible ray count.
%For out light sampling strategy we use a 2-level process, the first level being a coarse voxel grid around the current scene volume.
%The second level is then a finer set of resampling reservoirs \cite{10.1145/3478512.3488613} that sample from a corresponding 1st level voxel grid cell.
%This hierarchical separation allows for light sampling of larger light sets within a fixed sampling budget compared to just using a single-level strategy on its own.

To importance sample the light candidates, we use a form of ReSTIR but implement world-space reservoir reuse described by Boiss\'{e} instead of screen space \cite{10.1111:cgf.14378, 10.1145/3478512.3488613}.
Indeed, screen-space reuse is not amenable to sampling the lighting at secondary path vertices, as neighbor vertices in world space are not necessarily close, or even available, in screen space.
We start by generating a reservoir for each of the hit positions using Resampled Importance Sampling (RIS) \cite{10.2312:EGWR:EGSR05:139-146} and Weighted Reservoir Sampling (WRS) \cite{Chao1982WRS}.
We will describe in the next section how we try and select the most important lights for a given point to initialize the reservoirs better.

The generated reservoirs are then stored inside a hash grid, using a similar approach to the spatial hashing technique used for the world cache.
In this case, however, we only supply the adaptively quantized position of the shaded vertex as input to our hash functions.
Still, we store the surface normal into an additional data stream, which allows for bilateral thresholding during the reuse, helping to reduce the darkening bias introduced by ReSTIR \cite{10.1145/3478512.3488613}.

To improve performance for real-time applications, our approach deviates from a typical ReSTIR implementation and aims at reducing the number of shadow rays generated per reservoir and the number of resampling passes.
Our approach skips the visibility ray altogether and only uses a single shadow ray after the resampling has been completed.
In effect, this reduces the required number of shadow rays to one per reservoir, but it increases the bias in the form of darkening, as highlighted in Figure \ref{fig:restir}.
We also forego the spatial resampling passes and rely on a single temporal reuse pass.
This approach reduces the noise within a fixed raytracing budget, allowing the world cache to be less aggressive with how the radiance is retained temporally.
This is an essential requirement for achieving low latency response to lighting changes.

\subsubsection{Light Grid Lookup Structure}

Reservoir resampling helps improve the light sampling quality by effectively sharing the sampling effort across neighbor vertices.
However, resampling can only be as good as the initial state it is being fed.
Therefore, we look to improve this initial state by generating a global light grid structure storing the most important lights for each spatial cell prior to the reservoir generation.
This helps produce better candidates for the reservoir generation phase, therefore greatly improving on the resampled output produced by our reservoir pipeline.

We generate the light grid by first determining its bounds from the positions of every hit point.
From there, we can create an axis-aligned bounding box (AABB) around the scene extents containing all these intersection positions.
Within these bounds, we then create a uniform voxel grid.
Each grid axis dimension is calculated to ensure grid cells remain square and no single axis has more than a user-configurable maximum cell count.
Each grid cell finally holds a fixed-size list of the most important lights for that scene region.
The size of this list is determined to be the minimum between a user-configurable value and the number of lights in the scene.
To weight each light's importance to the grid cell, we calculate the relative luminance of the total deposited light from each light source over the cell's total volume. The total deposited light is calculated from the radiance \(L\) at each point \(p\) within the cells volume \(V\) using:

\begin{equation} \label{light_volume}
\iiint_V L(p) \,dx\,dy\,dz
\end{equation}

Equation (\ref{light_volume}) must be solved for all supported light types which are point, spot, directional, area, and environment in our implementation.
However, it is not possible to integrate this quantity analytically.
One numerical solution is to generate multiple samples within the volume and combine them in order to approximate the integral \cite{10.5555/1854996}.
This has the downside of requiring many samples to properly converge to the correct value and ensure all lighting is correctly detected when lights only partially cover the cell.
For instance, a spotlight whose cone only passes through a small volume region may be entirely skipped if none of the generated samples fall within the light cone.
Instead, we use a fast, cheap approximation to Equation (\ref{light_volume}) by sampling the light radiance at each corner of the grid cell and tri-linearly interpolating across the entire volume.
During this step, we ignore the cone angle of spotlights and treat them as point lights.
We then prepend an additional step to cull lights quickly by clamping the area of effect of each light. We then perform an intersection test between the light's area of effect and the grid cell and ignore lights that do not intersect.
For point and spot lights, this area of effect is based on the light’s user-supplied maximum distance attribute.
Area lights, however, have an infinite area of effect, so we calculate a maximum distance value by determining the range in which the light's contribution falls below a user-supplied threshold value.
This threshold value can thus be used to trade performance for light sampling bias.
For point lights, the area of effect becomes a sphere; for spotlights, it is their outer cone; and for area lights, it is the hemisphere above the surface.
We further optimize this test by treating each cell's AABB as a bounding sphere, as this significantly simplifies the calculations while only introducing minimal error.
The above algorithm trades accuracy for performance in multiple locations, but these inaccuracies are hidden by the subsequent reservoir resampling passes.

The light grid structure is finally used to feed samples as batches into each of the generated reservoirs.
Samples are selected by generating a random index into the light list of the grid cell overlapping the shaded vertex and are then resampled into a reservoir using the procedure described in section \ref{subsec:restir}.

\subsection{Irradiance Estimation}
\label{subsec:irradiance_estimation}

We have seen that our screen cache encodes the incoming radiance in every direction over the primary visible surfaces.
However, we are still left to estimate the irradiance.
This section  describes how to leverage our probe grid and evaluate the irradiance for every pixel on the screen.

\subsubsection{Per-Pixel Interpolation}
\label{subsec:probe_interpolation}

As probes are placed sparsely across the screen, we compute a weighted average of the radiance from the neighbor probes to reconstruct the lighting signal.
This process, usually called interpolation, is performed for every pixel at the target resolution.
If we reflect on the rest of the pipeline presented so far, this is the only pass running at full rate, with regards to target resolution, along with the probe reprojection pass described in \ref{subsec:temporal_upscale}.

\begin{figure}
  \centering
  \includegraphics[width=0.97\linewidth]{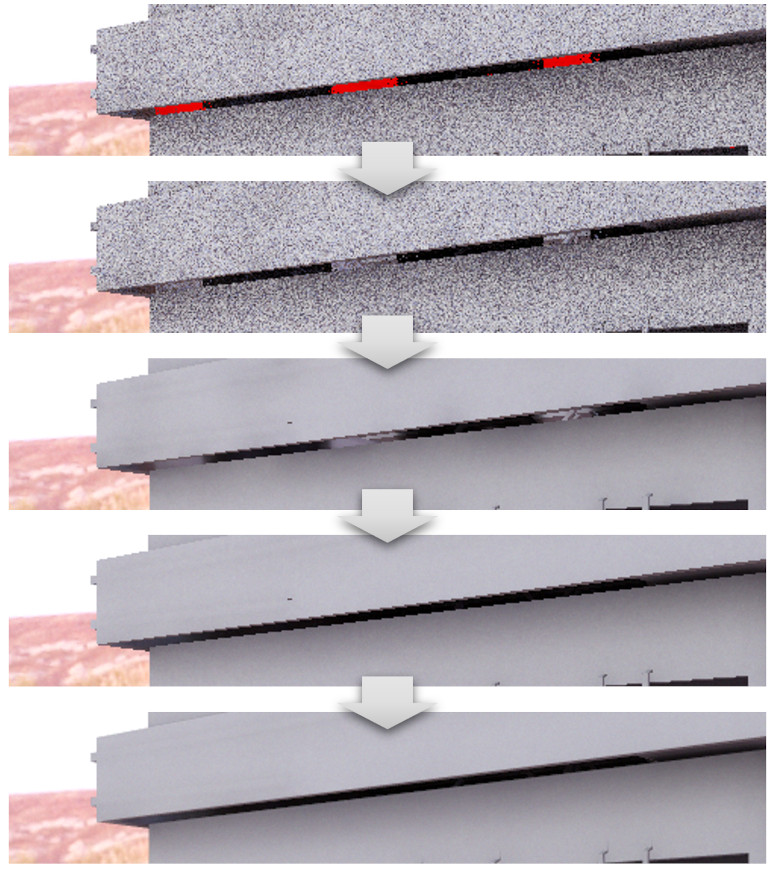}
  \caption{Fixing interpolation failures using denoiser hint.}
  \label{fig:denoiser-hint}
\end{figure}

We start out by finding the neighboring probes in all 4 directions using the $find\_closest\_probe()$ function defined in \ref{subsec:probe_masking}.
The contribution for each probe is weighted using an edge-aware function based on the surface depth and normal.
In this process, we also ensure that nearby probes, lying further away than $cell\_size$ units from the pixel's plane, are discarded by assigning a weight of $0$.
If all probes get assigned a null weight, then interpolation fails, which can lead to light-leaking artifacts as highlighted in Figure \ref{fig:denoiser-hint}.
In such cases, we fall back to setting equal weights for all neighbor probes, which we refer to as "relaxed interpolation".
We flag pixels calculated using relaxed interpolation in the alpha channel of the resolved texture; this information will be used later as a hint for the denoiser to try and discard the evaluated sample.

Finally, we jitter the pixel's position before performing the search for neighbor probes to break up the structured artifacts resulting from all neighboring pixels interpolating from the exact same set of probes.
It is worth noting, however, that we cancel the jitter if the resulting position lands outside the original pixel's plane \cite{lumen}.
This cancellation is necessary to prevent an increase in the number of pixels relying on relaxed interpolation.

\begin{figure}
  \centering
  \begin{subfigure}{0.15\textwidth}
    \centering
    \includegraphics[width=\linewidth]{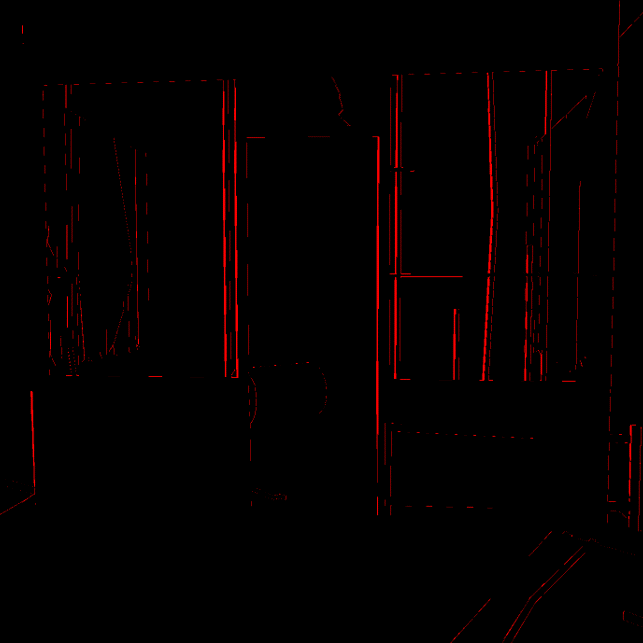}
    \caption{Disocclusion mask}
  \end{subfigure}
  \begin{subfigure}{0.15\textwidth}
    \centering
    \includegraphics[width=\linewidth]{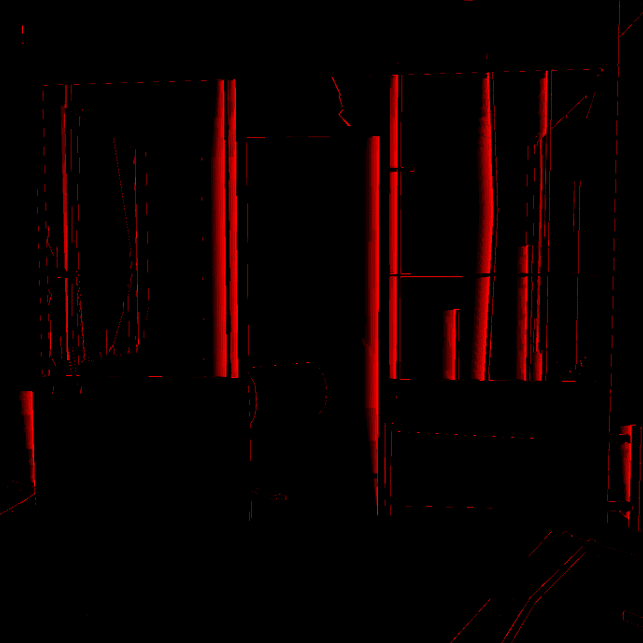}
    \caption{Dilated blur mask}
  \end{subfigure}
  \begin{subfigure}{0.15\textwidth}
    \centering
    \includegraphics[width=\linewidth]{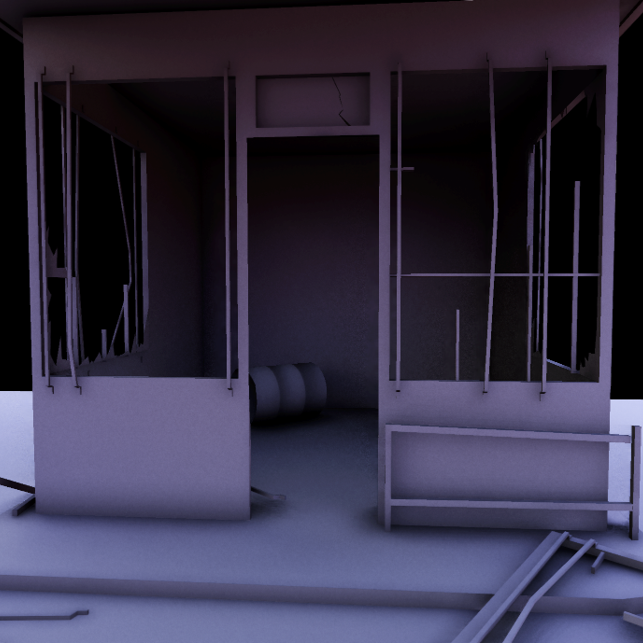}
    \caption{Filtered irradiance}
  \end{subfigure}
  \caption{Spatial filtering guided by dilated blur mask.}
  \label{fig:blur_mask}
\end{figure}

\subsubsection{Spherical Harmonics}

To accurately evaluate the irradiance encoded by a given probe, we need to fetch nearly all the stored radiance samples, that is, all the mapped directions across the oriented hemisphere.
We instead project the screen probes to spherical harmonics (SH) prior to the interpolation pass.

Spherical harmonics indeed provide several key benefits to estimating the irradiance:
\begin{itemize}
    \item High-frequency noise can be filtered at no additional cost by limiting the projection to the first three bands.
    \item Irradiance can be precisely estimated while reducing the requirements on the amount of memory to be fetched.
    \item Finally, we update our reprojection pass also to reproject the SH representation of each probe. We, therefore, only need to perform the projection for the newly spawned probes, which drastically reduces the cost of the operation.
\end{itemize}

The irradiance is then computed using the dot product of the projected cosine lobe with the projected radiance.
The projected radiance is fetched from the cache while the cosine lobe is projected analytically \cite{10.1145/383259.383317}.

\subsubsection{Denoising}
\label{subsec:probe_interpolation_denoising}

The interpolated irradiance typically exhibits low-frequency noise from the probes and high-frequency noise caused by the pixel jittering mentioned in \ref{subsec:probe_interpolation}.
The low-frequency noise manifests itself as boiling in areas where the incoming radiance is the most noisy, while the high-frequency noise is similar to that of a dithering pattern.
These issues can be solved using a simple denoiser based on temporal accumulation and an adaptive spatial filter, where we compute the spatial filter radius depending on the number of samples accumulated in history.
As not all pixels have a history, such as disoccluded pixels, we adapt the filtering radius to the number of accumulated samples to reduce the noise as illustrated in Figure \ref{fig:blur_mask}.
%Disoccluded areas will typically use a large radius. and areas with an old history no longer have spatial filter, as illustrated in Figure \ref{fig:blur_mask}.
The spatial filter uses an edge-aware weight based on depth and normal.
When interpolation fails and no history is available, we use the relaxed irradiance interpolation instead of outputting no irradiance.
% TODO: how we dilate blur mask and why, don't remember the why...

\section{Screen-Space Global Illumination}
\label{sec:ssgi}

The technique described so far can typically lack details as small-scale occlusion and color bleeding, smaller than the space between screen probes, cannot be captured by the grid.
We propose a hybrid approach to mitigate this issue and implement short-range screen-space ambient occlusion and global illumination to fill in the missing details.
This section explains how we combine screen-space global illumination with the pipeline we have described so far.

\begin{figure}
  \centering
  \begin{subfigure}{0.23\textwidth}
    \centering
    \includegraphics[width=\linewidth]{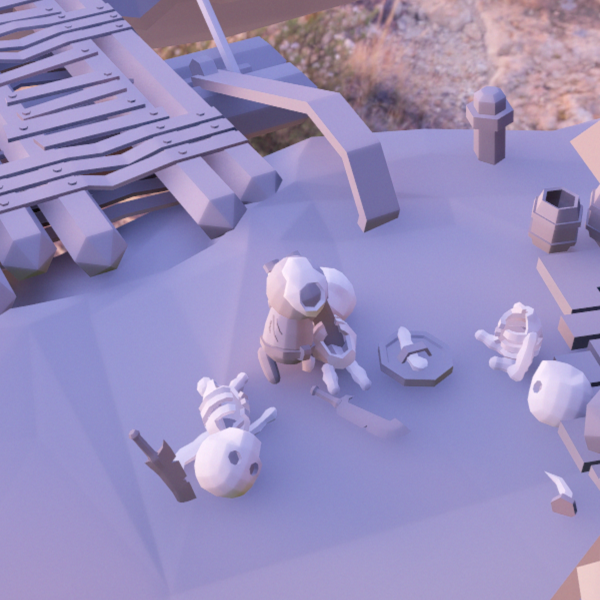}
    \caption{GI-1.0 irradiance estimation}
  \end{subfigure}
  \begin{subfigure}{0.23\textwidth}
    \centering
    \includegraphics[width=\linewidth]{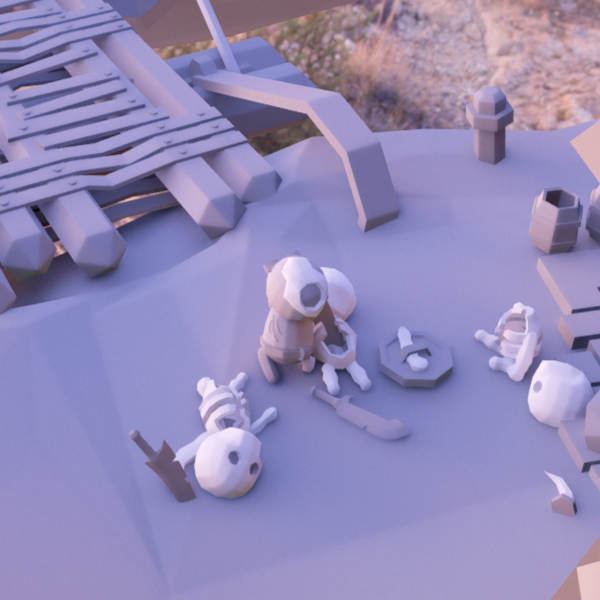}
    \caption{Combined w/ screen-space GI}
  \end{subfigure}
  \caption{Using short screen traces to recover small details.}
  \label{fig:ssgi}
\end{figure}

\subsection{Combining Near and Far Fields}

As the screen cache captures the low-frequency global illumination well, we only add global illumination from screen-space techniques where the high-frequency information is missing.
Following \cite{Jimenez2016PracticalRS} and \cite{Mayaux2018HBIL}, we create a mask using a bent cone, which is a cone centered on the bent normal with an aperture angle derived from the ambient occlusion.
Instead of integrating the clamped cosine lobe with the incoming radiance, we multiply the bent cone and the clamped cosine lobe, then integrate the product with the incoming radiance.
Ringing artifacts can occur in darker areas.
This issue can be mitigated by using windowing with a small factor, as done in \cite{shtricks}.

\subsection{Horizon-Based Occlusion}

While most screen-space methods could be used or extended, we focus on horizon-based methods \cite{Jimenez2016PracticalRS} \cite{Mayaux2018HBIL}, which offer a good trade-off between performance and quality for global near-field illumination (i.e., using a short radius).
We use temporal accumulation for removing interpolation artifacts after the irradiance integration, so we only compute one slice per frame using a few steps.
The bent normal and ambient occlusion estimation leads to a noisy mask approximation, which we denoise using the spatiotemporal loop described in \ref{subsec:probe_interpolation_denoising}.

\section{Ray Traversal}

The method described in this paper requires ray casting operations which we implement with hardware-accelerated ray tracing using a traditional Bounding Volume Hierarchy (BVH) to accelerate intersection calculations.
However, the global illumination algorithm can use other methods for ray casting as long as the method returns the hit point of a ray.
We also implemented ray casting using a hybrid of hardware-accelerated BVH ray tracing and distance field tracing \cite{cite:pjhybrid}.
We improved the ray tracing quality by using BVH ray tracing at the place where distance field tracing only produces a large error.
Although it was implemented, the speed-up we observed was not big enough to employ it as the default option, as our technique is already optimized to trace a reduced number of rays.
The method would pay off if the algorithm requires more ray casting. 

\section{Implementation and Results}

\begin{figure*}
  \centering
  \begin{subfigure}{0.475\textwidth}
    \centering
    \includegraphics[width=\linewidth]{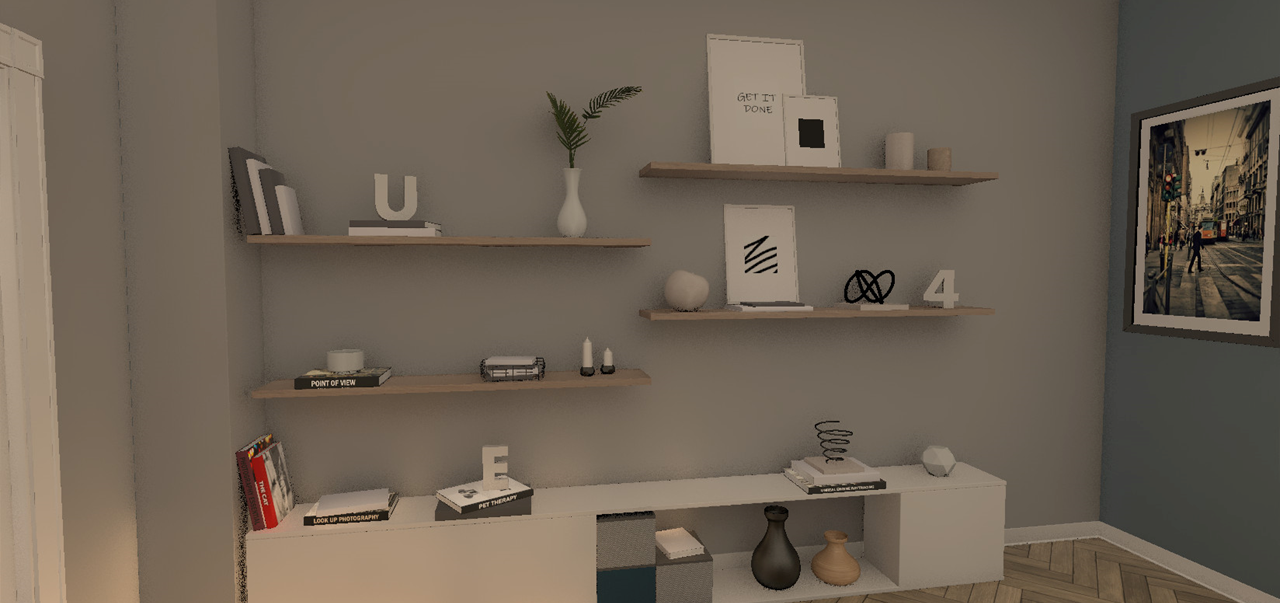}
    \caption{Lumen's output using SW traversal (2.6ms)}
  \end{subfigure}
  \begin{subfigure}{0.475\textwidth}
    \centering
    \includegraphics[width=\linewidth]{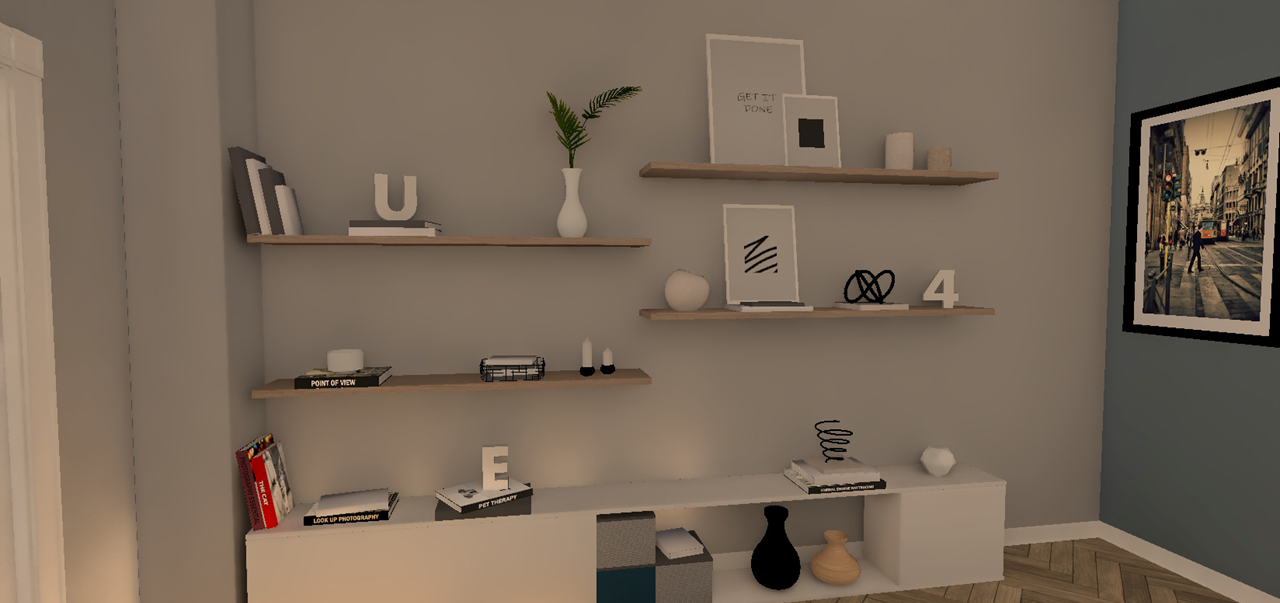}
    \caption{GI-1.0's output using HW traversal (2.4ms)}
  \end{subfigure}
  \caption{Running our GI-1.0 pipeline inside Unreal Engine 5.}
  \label{fig:lumen}
\end{figure*}

\begin{table}
    \centering
    \footnotesize
    \caption{GI-1.0 performance results (in ms)}
    \label{tab:results}
    \begin{center}
    \setlength\tabcolsep{0.2pt}
    \begin{tabular}{|c|c c|c c|c c|c c|}
        \hline
        & \multicolumn{2}{c }{\begin{tabular}{c} {Raytracing}     \\ {(DXR-1.1)}     \end{tabular}}
        & \multicolumn{2}{c }{\begin{tabular}{c} {Caching}        \\ {\& sampling}   \end{tabular}}
        & \multicolumn{2}{c }{\begin{tabular}{c} {Interpolate}    \\ {\& denoise}    \end{tabular}}
        & \multicolumn{2}{c|}{\begin{tabular}{c} {\textbf{Total}} \\ {\textbf{time}} \end{tabular}} \\
        \hline\hline
        & \begin{tabular}{c} {\tiny AMD} \\ {\tiny Radeon\texttrademark} \\ {\tiny RX 6900 XT} \end{tabular} & \begin{tabular}{c} {\tiny NVIDIA} \\ {\tiny GeForce} \\ {\tiny RTX\texttrademark{} 3080} \end{tabular}
        & \begin{tabular}{c} {\tiny AMD} \\ {\tiny Radeon\texttrademark} \\ {\tiny RX 6900 XT} \end{tabular} & \begin{tabular}{c} {\tiny NVIDIA} \\ {\tiny GeForce} \\ {\tiny RTX\texttrademark{} 3080} \end{tabular}
        & \begin{tabular}{c} {\tiny AMD} \\ {\tiny Radeon\texttrademark} \\ {\tiny RX 6900 XT} \end{tabular} & \begin{tabular}{c} {\tiny NVIDIA} \\ {\tiny GeForce} \\ {\tiny RTX\texttrademark{} 3080} \end{tabular}
        & \begin{tabular}{c} {\tiny AMD} \\ {\tiny Radeon\texttrademark} \\ {\tiny RX 6900 XT} \end{tabular} & \begin{tabular}{c} {\tiny NVIDIA} \\ {\tiny GeForce} \\ {\tiny RTX\texttrademark{} 3080} \end{tabular} \\
        \hline
        \begin{tabular}{c} {Gas station} \\ {\scriptsize{(99k triangles)}} \\ {\includegraphics[width=0.16\linewidth]{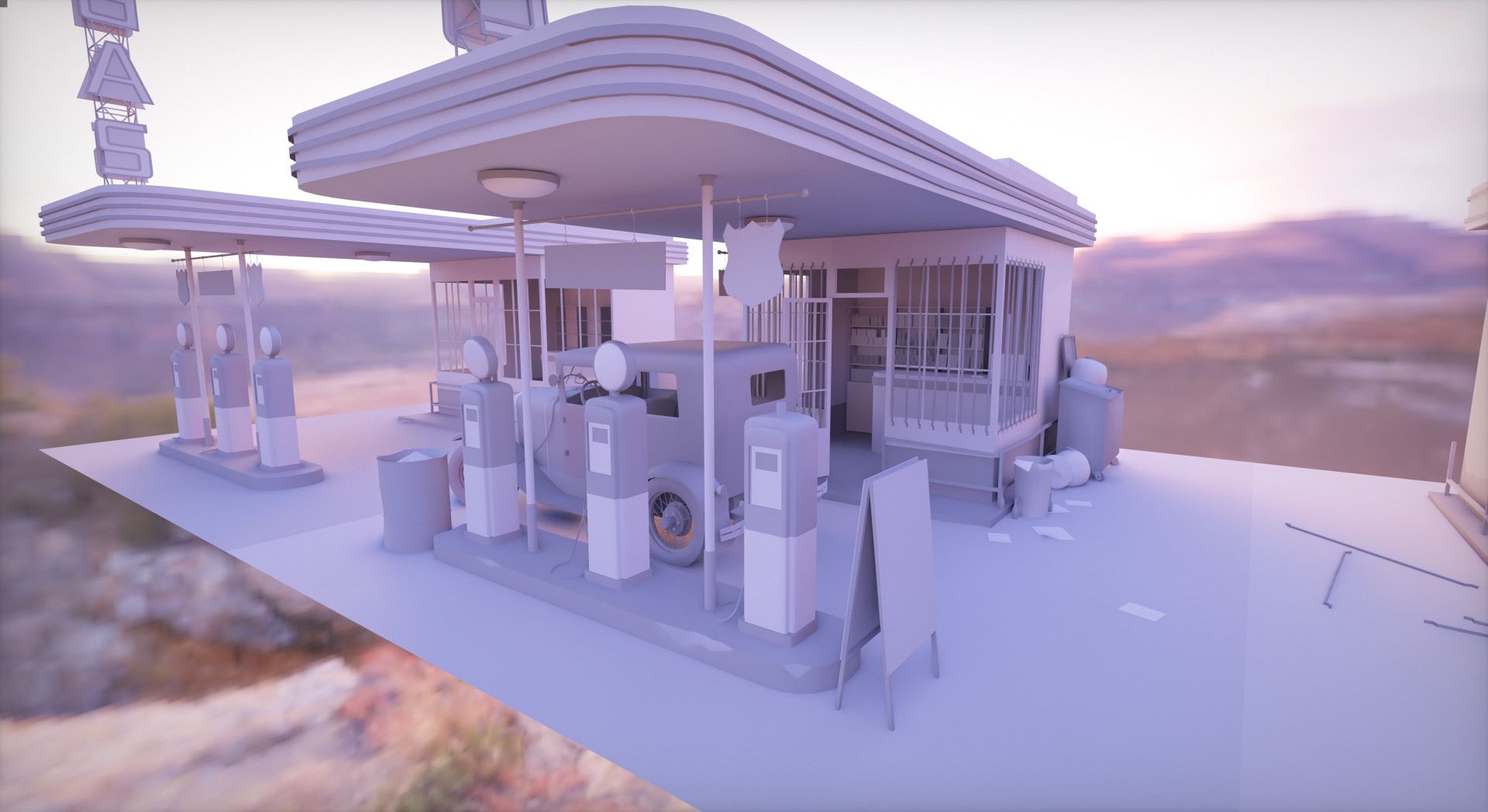}} \end{tabular}
        & 0.235 & 0.132 & 1.032 & 1.220 & 0.665 & 0.554 & \textbf{1.932} & \textbf{1.906} \\
        \hline
        \begin{tabular}{c} {Flying world} \\ {\scriptsize{(267k triangles)}} \\ {\includegraphics[width=0.16\linewidth]{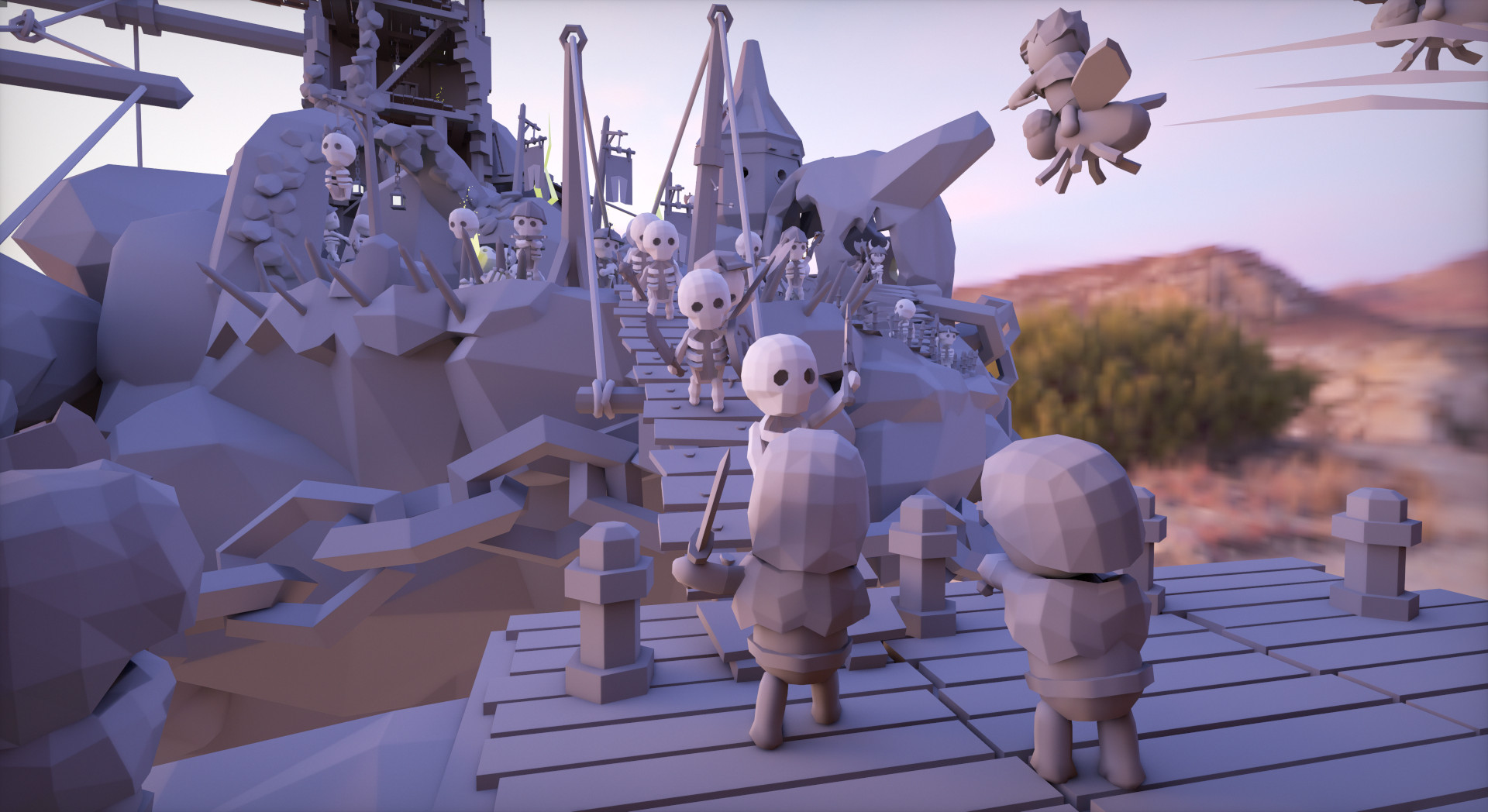}} \end{tabular}
        & 0.532 & 0.292 & 1.226 & 1.460 & 0.705 & 0.574 & \textbf{2.463} & \textbf{2.326} \\
        \hline
        \begin{tabular}{c} {Kitchen \#1} \\ {\scriptsize{(1.4M triangles)}} \\ {\includegraphics[width=0.16\linewidth]{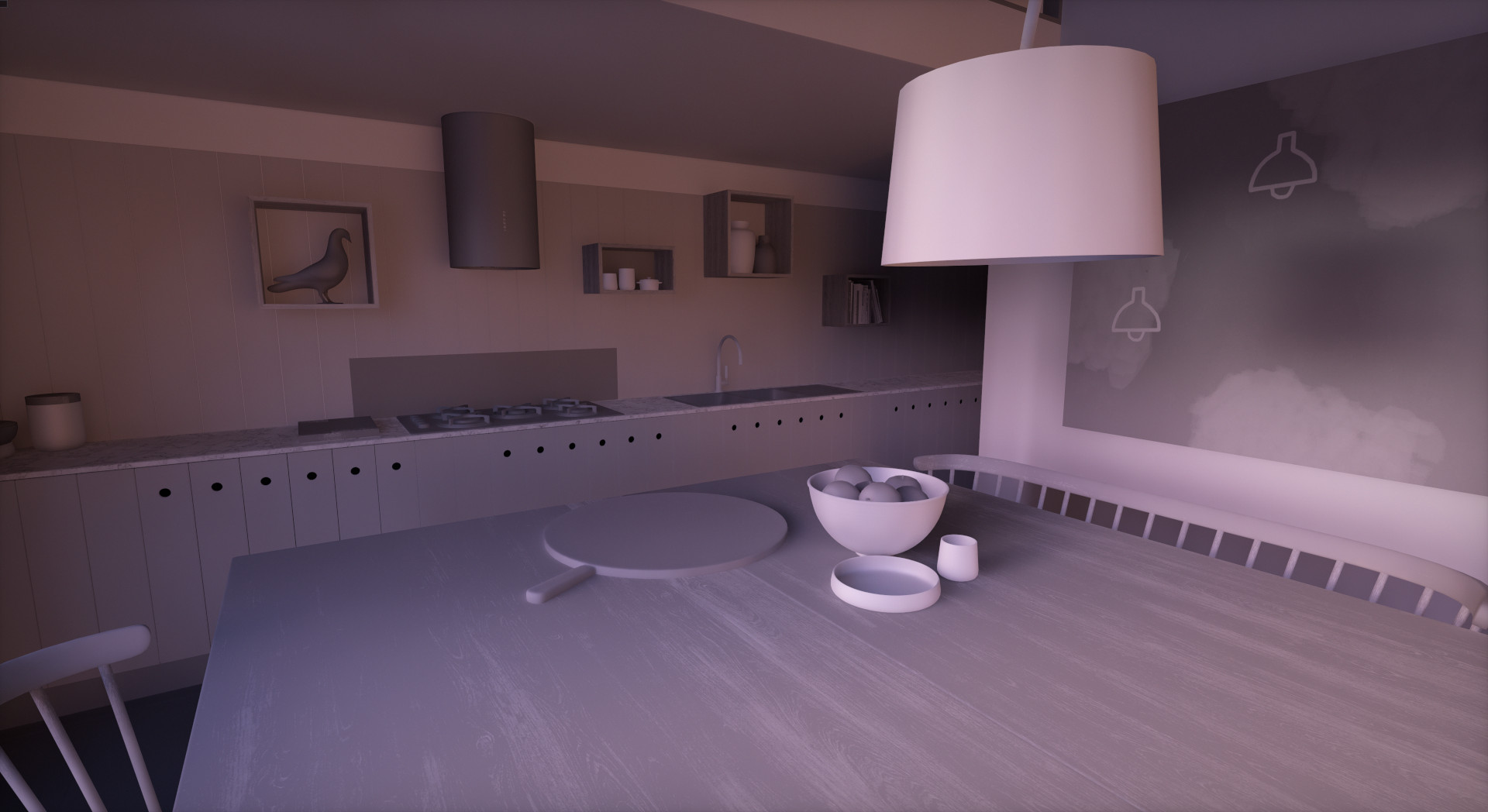}} \end{tabular}
        & 0.336 & 0.198 & 1.633 & 2.122 & 0.765 & 0.656 & \textbf{2.734} & \textbf{2.976} \\
        \hline
        \begin{tabular}{c} {Kitchen \#2} \\ {\scriptsize{(9.0M triangles)}} \\ {\includegraphics[width=0.16\linewidth]{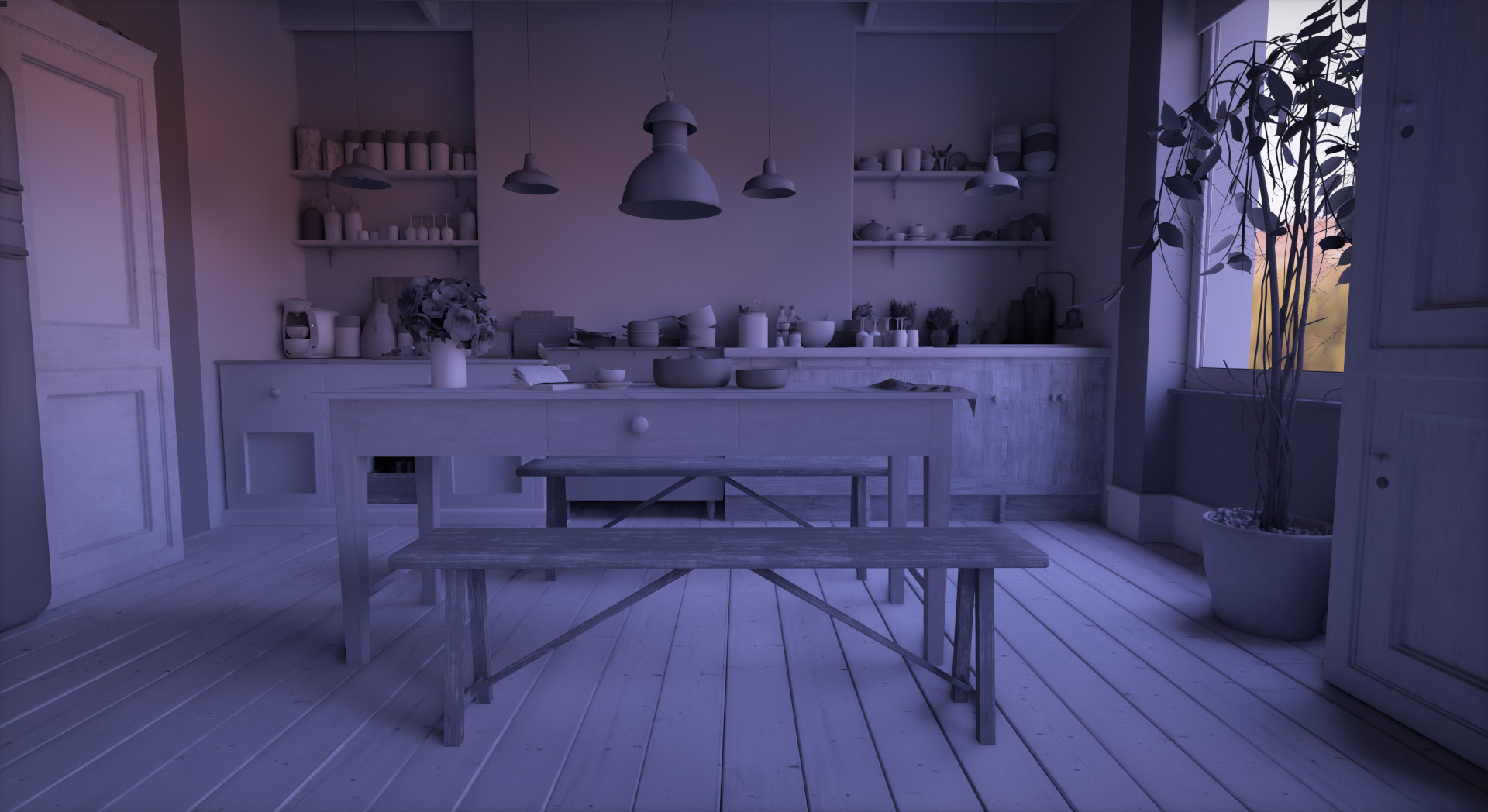}} \end{tabular}
        & 0.676 & 0.343 & 1.693 & 2.327 & 0.755 & 0.756 & \textbf{3.124} & \textbf{3.426} \\
        \hline
    \end{tabular}
    \end{center}
\end{table}

\begin{figure*}
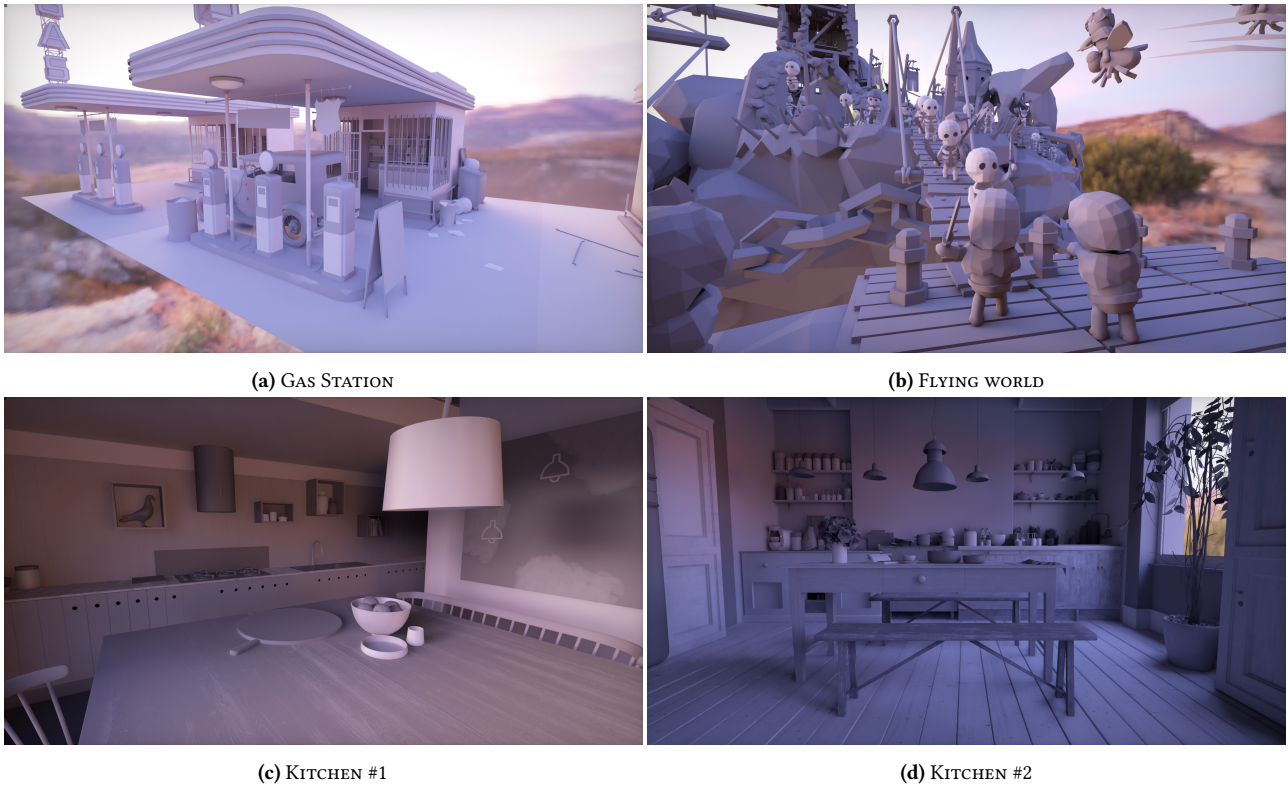

    \begin{subfigure}[b]{1.0\columnwidth}
        \centering
        \includegraphics[width=\textwidth]{tab/results/gas_station.jpg}
        \caption{{\sc Gas Station}}
    \end{subfigure}
    \begin{subfigure}[b]{1.0\columnwidth}
        \centering
        \includegraphics[width=\textwidth]{tab/results/skeletons.jpg}
        \caption{{\sc Flying world}}
    \end{subfigure}
    \begin{subfigure}[b]{1.0\columnwidth}
        \centering
        \includegraphics[width=\textwidth]{tab/results/kitchen.jpg}
        \caption{{\sc Kitchen \#1}}
    \end{subfigure}
    \begin{subfigure}[b]{1.0\columnwidth}
        \centering
        \includegraphics[width=\textwidth]{tab/results/big_kitchen.jpg}
        \caption{{\sc Kitchen \#2}}
    \end{subfigure}
    \caption{Test scenes. }
    \label{fig:results}
\end{figure*}

We implemented our GI-1.0 pipeline inside our Direct3D12 research framework using DXR-1.1 for accelerating the ray intersection queries. 
We prepared some test scenes shown in Figure \ref{fig:results} and used them to evaluate the runtime performance of our technique, estimating both the direct and indirect lighting at \textonequarter{} samples per pixel at 1080p. 
%mention something about FSR?

We have broken down the timings into 3 categories:
\begin{itemize}
    \item \emph{Raytracing} represents approximately the time spent in ray traversal.
    We mention approximately here as other bits of logic are run in both the closest-hit kernel (i.e., screen cache rays) and any-hit kernel (i.e., world cache rays). However, the cost is vastly dominated by ray traversal.
    \item \emph{Caching \& sampling} represents the time spent maintaining the screen cache and world cache representations.
    This includes all guiding, sampling, reconstruction, filtering, prefiltering, reprojection, etc.
    \item \emph{Interpolate \& denoise} represents the time spent in the per-pixel interpolation pass followed by the spatiotemporal denoiser.
    This includes projecting the screen probes to spherical harmonics and estimating the disocclusion mask, further dilated into a blur mask for filtering.
\end{itemize}

Table \ref{tab:results} shows the breakdown and the total time spent executing our GI-1.0 pipeline \footnote{Based on AMD internal testing, September 2022, using a desktop system configured with a Radeon\texttrademark\ RX 6900 XT GPU, Ryzen\texttrademark\ 7 5800X CPU, 32GB RAM, and Windows 10 vs. a similarly configured system with an NVIDIA RTX\texttrademark\ 3080 GPU to measure the performance results (in ms) of both systems in DXR-1.1 raytracing, caching \& sampling, and interpolation \& denoising. Results will vary.}.
We can see that the total time ranges from 1.932ms to 3.124ms on an AMD Radeon\texttrademark\ RX 6900 XT GPU.
We achieve this performance with small amounts of noise and at a low raytracing cost thanks to our caching scheme.
A recorded video of our technique is also available at GPUOpen.com. 

We also implemented our global illumination pipeline as a plugin for Unreal Engine 5 (UE5).
This allowed us to validate our technique and choices across a wider variety of environments and lighting scenarios.
It further enabled the ability to switch between our solution and UE5's Lumen renderer for performance and quality comparisons.
We present this comparison, considering only the indirect lighting signal, using the sample UE5 Archviz scene in Figure \ref{fig:lumen}.
These images are screen captures of the UE5 engine running Lumen and GI-1.0. 

It is worth noting, however, that our UE5 integration remains incomplete, resulting in materials being approximated in our plugin.
This is not a limitation of our GI-1.0 pipeline but a consequence of the important engineering effort required to evaluate UE5's material system accurately.
%, which we deemed not worth spending much time on.

\section{Limitations and Future Work}

We have presented a complete raytraced global illumination pipeline aimed primarily at estimating the indirect diffuse lighting of a scene dynamically at runtime.
Direct lighting from environment maps and emissive surfaces is also supported, at no additional cost, thanks to the robustness of our ray guiding implementation.

\subsection{Limitations}

We have mentioned in section \ref{subsec:radiance_feedback} that we could estimate an infinite number of light bounces over multiple frames thanks to our radiance feedback mechanism.
While this technique provides some interesting performance advantages, on top of the visual improvements, it only really works when the contributing reflector is visible inside the previous frame.
This can be detrimental to visual fidelity, particularly in interior scenes, where bounced lighting dominates.
In some future work, we want to look at efficient ways of directly approximating path continuation  in hash space to achieve fast and reliable multi-bounced lighting everywhere in world space.

\subsection{Glossy Reflections}

Glossy reflections can be rendered with stochastic methods similar to \cite{cite:stachowiak}.
Although our implementation of glossy reflections is still in progress, here we briefly describe the algorithm we are implementing.
We generate importance-sampled directions from the GGX lobe \cite{Heitz2018GGX} and evaluate the reflected radiance using our screen and world caches.
As screen probes store low-frequency illumination, directly using this representation for low-roughness surfaces gives an overly blurry result.
In such situations, we instead cast a ray in the sampled direction and evaluate the incoming radiance by accessing the environment map if the ray misses or the world cache at the hit position.
One advantage of using our caching hierarchy for reflections is that we can account for multi-bounce lighting solely by looking up our cached radiance in memory.

\subsection{Future Work}

The current light grid sampling approach provides a list of important lights for a given scene area; it is interesting to note that this also includes the visible region of the view frustum.
As such, it may be helpful to leverage this structure to sample the lighting at primary path vertices and enable general direct lighting support from many light sources.
The metric used in calculating the light grid importance adds bias to the rendering as we clamp the light intensity.
Adding a non-biased approach is future work \cite{slc}.

\begin{acks}
    We would like to thank Bruno Stefanizzi and Prashanth Kannan for their support of the research, as well as Holger Gruen and Oleksandr Kupriyanchuk for their help in reviewing this paper. 
    \href{https://sketchfab.com/3d-models/gas-stations-fixed-b6e9be9f475a4930865e281e02539dc6}{Gas station} and \href{https://sketchfab.com/3d-models/flying-world-battle-of-the-trash-god-350a9b2fac4c4430b883898e7d3c431f}{Flying world} were created by John Constantine and burunduk, respectively.
    AMD, AMD Radeon and the AMD Arrow logo, and combinations thereof are trademarks of Advanced Micro Devices, Inc.  Other product names used in this publication are for identification purposes only and may be trademarks of their respective companies.
\end{acks}

\bibliographystyle{ACM-Reference-Format}
\bibliography{main}
\end{document}